\documentclass[final,5p,]{elsarticle}

\usepackage[american]{babel}
\usepackage{amssymb,amsmath}
\usepackage[utf8]{inputenc}
\allowdisplaybreaks[3]
\usepackage{graphicx}
\usepackage{color}
\usepackage{xspace}
\usepackage{textcomp}
\usepackage{verbatim}
\usepackage{hyperref}
\usepackage{microtype}
\usepackage[capitalise]{cleveref}
\usepackage{pgfplots}
\usepgfplotslibrary{colormaps}
\pgfplotsset{compat=1.13}
\usepackage{onimage}
\usepackage{tikz}
\usetikzlibrary{shapes,arrows.meta,math}
\usepackage{algcompatible}
\usepackage{float}
\newfloat{algorithm}{tbhp}{loa}
\floatname{algorithm}{Algorithm}

\usepackage[firstpage=true,color=black,placement=top,scale=1,opacity=1,vshift=-1cm]{background}

\SetBgContents{
\begin{minipage}{2\linewidth}
\small
\copyright 2019. This manuscript version is made available under the CC-BY-NC-ND 4.0 license \\\url{http://creativecommons.org/licenses/by-nc-nd/4.0/} \\
A definitive version was published in \textbf{Ultramicroscopy, Volume 198, March 2019, Pages 49--57} and is available at \url{https://dx.doi.org/10.1016/j.ultramic.2018.12.016}
\end{minipage}
}

\providecommand{\abs}[1]{\left\lvert#1\right\rvert}
\providecommand{\Abs}[1]{\left\lVert#1\right\rVert}
\providecommand{\brac}[1]{\left(#1\right)}

\DeclareMathOperator{\dx}{d\mathit{x}}
\DeclareMathOperator{\identityMapping}{id}
\DeclareMathOperator{\identity}{1\!\!1}

\DeclareGraphicsExtensions{.pdf,.jpg,.png}

\graphicspath{
{./Figs/}
}

\newlength{\imgwidth}

\journal{Ultramicroscopy}

\begin{document}

\begin{frontmatter}

\title{Joint non-rigid image registration and reconstruction for quantitative atomic resolution scanning transmission electron microscopy}

\author[1]{Benjamin Berkels}
\author[2]{Christian H. Liebscher}
\address[1]{AICES, RWTH Aachen University, Aachen, NRW, Germany}
\address[2]{Max-Planck-Institut für Eisenforschung, Düsseldorf, NRW, Germany}

\begin{abstract}
Aberration corrected scanning transmission electron microscopes (STEM) enable to determine local strain fields, composition and bonding states at atomic resolution. The precision to locate atomic columns is often obstructed by scan artifacts limiting the quantitative interpretation of STEM datasets. Here, a novel bias-corrected non-rigid registration approach is presented that compensates for fast and slow scan artifacts in STEM image series. The bias-correction is responsible for the correction of the slow scan artifacts and based on a explicit coupling of the deformations of the individual images in a series via a minimization of the average deformation. This allows to reduce fast scan noise in an image series and slow scan distortions simultaneously. The novel approach is tested on synthetic and experimental images and its implication on atomic resolution strain and elemental mapping is discussed.
\end{abstract}


\end{frontmatter}

\section{Introduction}
Since the invention of aberration correction, scanning transmission electron microscopy (STEM) has become a powerful technique to image and analyze materials at ultimate atomic resolution \cite{Pennycook2011}. The correction of lens errors enables outstanding resolution of down to 50 pm and a multitude of signals can be collected in each probe position simultaneously, since the convergent electron beam is scanned over the sample \cite{Krivanek2015}. With the advancement of X-ray spectrometers for example, it is nowadays even possible to quantitatively determine the elemental distribution in a sample at atomic resolution \cite{Kothleitner2014}. A crucial aspect not only for spectroscopic imaging, but quantitative atomic resolution STEM in general, is the instrumental stability and with this the degree of scan noise and artifacts in an image \cite{BrBoLa2012,JoNe13}, as well as sample drift \cite{KiAsYu10}.

Novel ways to identify and compensate scan noise and linear sample drift of individually recorded STEM images were introduced by Jones and Nellist~\cite{JoNe13}. In an early study, Kimoto et al.~\cite{KiAsYu10} applied a rigid alignment of multiple images acquired with high pixel dwell time to improve the signal to noise ratio in atomic resolution STEM images. An approach to account for non-linear sample drift was introduced as revolving STEM (RevSTEM) \cite{SaLe14}. Here, each image in an image series with high frame rate is rotated to determine the sample drift vector. Ophus et al.~\cite{Ophus2016a} developed a technique where only two orthogonal scans at medium frame rates are required. The non-linear distortion introduced by sample drift is reduced by correcting the scan line origins. A similar approach was introduced recently by Ning et al.~\cite{Ning2018} to also reduce STEM noise from orthogonally scanned image pairs. Novel approaches even apply complex scan patterns, such as spiral beam paths, to further reduce scan distortions and avoid line-by-line distortions stemming from the fast and slow scan directions of conventional scan patterns~\cite{Sang2017}.

High frequency instabilities of $>$50 Hz, typically resulting from electronic noise, influence the scan system of the microscope and can effectively be reduced by employing a serial image acquisition scheme \cite{Berkels2012}. Here, images acquired at high frame rates are aligned and averaged to mainly compensate for fast scan noise. Accounting for local image distortions in consecutive frames led to the development of non-rigid image registration techniques \cite{BeBiBl13,JoYaPe15}. It was shown that even sub-picometer precision can be obtained in locating atomic column positions with increased signal-to-noise ratio \cite{YaBeDa14}. Jones at el.\ \cite{Jones2018} also showed that optimizing the acquisition conditions of the serial imaging and utilizing non-rigid registration greatly reduces the influence of slow scan distortions on atomically resolved strain maps. Correcting local image deformations is not only crucial to increase the precision in locating atomic column positions, but is also applicable to multi-frame spectroscopic datasets \cite{Jones2018,YaZhOh16,Wang2018}.

Here, we report about a novel extension of the non-rigid registration algorithm developed in \cite{BeBiBl13} and its application to synthetic and experimental datasets to effectively reduce fast and slow scan noise in atomically resolved STEM images. In the first part, the mathematical description of the bias-corrected non-rigid registration approach is presented and tested on a synthetic dataset with intentionally introduced fast and slow scan noise. Bias-correction here means that the determined deformations of images in a series are no longer depending on the coordinate system of the first image, but rather on the average coordinate system of the input series. This requires a direct coupling of the deformations with each other in the optimization algorithm.
Methodologically, the main contribution of this paper is the addition of a deformation reduction step to the alternating minimization algorithm used in \cite{BeBiBl13} that introduces the direct coupling needed for the bias-correction.
In the subsequent part of the paper, the application of the new approach is tested on two experimental datasets acquired under different imaging conditions. The first experimental dataset shows the impact of the new bias-corrected approach on atomically resolved strain maps across a coherent interface. In the second example, its applicability and influence on a multi-frame acquisition under spectroscopic measurement conditions is demonstrated.

\section{Materials and methods}
The novel extension of the non-rigid registration algorithm is tested on two different material systems, a nano-dispersion hardened low-density steel and a C14-Fe$_2$Nb Laves phase. Details on the processing conditions of the materials can be found in \cite{Yao2016} and \cite{Voss2010}, respectively. The sample preparation routine by focused ion beam (FIB) milling of a needle-shaped specimen of the low-density steel is described in detail in \cite{LiYaDe18}. Samples of the C14-Fe$_2$Nb Laves phase with 3 mm in diameter and 500 $\mu$m thickness were obtained by electrical discharge machining. After mechanical polishing to 150 $\mu$m thickness, the foils were twin-jet polished to perforation using a Struers Tenupol-5 and an electrolyte of 90 vol.\% ethanol and 10 vol.\% perchloric acid at -30$^{\circ}$C.

STEM images were taken in a C$_{s}$ probe corrected Titan Themis 60-300 (FEI, Thermo Fisher Scientific) instrument with probe semi-convergence angles between 17 and 23.8 mrad and probe currents between 50 and 80 pA. The inner and outer collection angles of the annular detector were set from 58-75 mrad and 285-350 mrad, respectively. The image series for atomic resolution strain mapping in the nano-dispersion hardened steel were obtained for pixel dwell times of 2 $\mu$s with 20 to 30 images per series. The series acquired in the Laves phase contains 240 images taken with a pixel dwell times of 10 $\mu$s. The detailed parameters will be given in the text for each dataset.

\subsection{Bias-corrected averaging using non-rigid registration}
Consecutive images are aligned and averaged to account for sample drift and local distortions using the non-rigid registration strategy from \cite{BeBiBl13}, but with one novel extension that was first mentioned in \cite{LiYaDe18}.
In order to understand the need for an extension of \cite{BeBiBl13}, we first give a brief summary of this approach. Given a series of $n$ consecutively acquired images $f_1,\ldots,f_n$, the goal is to bring all images into a common coordinate system so that the images can be averaged into a single image with an improved signal-to-noise ratio. Given a candidate for the averaged image $f$ (starting with $f_1$ as initial guess), for each image $f_i$, one computes a deformation of the image domain $\phi_i$ such that $f_i\circ\phi_i\approx f$. Note that computing these deformations is numerically challenging due to the strong non-convexity of the associated optimization problems and requires sophisticated algorithms. Nevertheless, \cite{BeBiBl13} provides an algorithm for this and we refer the reader to this paper for further details on computing the $\phi_i$ for a given average image $f$. Given deformations $\phi_i$, the average image can then be computed by simply averaging (e.g. with mean or median) the deformed images $f_i\circ\phi_i$ pixel-wise. With the new average, one can again compute new deformations and iterate this procedure.
Essentially, this means that the objective functional
\begin{align*}
E[f,\phi_1,\ldots,\phi_n]=\sum_{i=1}^n\bigg(&-\text{NCC}[f,f_i\circ\phi_i]\\
&+\frac{\lambda}{2}\int_\Omega\Abs{D\phi_i-\identity}^2\dx\bigg)
\end{align*}
is minimized alternatingly with respect to the average $f$ and the deformations $(\phi_1,\ldots,\phi_n)$, i.e.\ one alternates between minimizing $E$ with respect to $(\phi_1,\ldots,\phi_n)$ for a fixed $f$ and minimizing $E$ with respect to $f$ for fixed $(\phi_1,\ldots,\phi_n)$.
One step of the alternating minimization strategy, i.e.\ first optimizing the deformations for a fixed average and then optimizing the average for fixed deformations, is called an outer iteration. The total number of outer iterations is a parameter of the method and denoted by $K$. Note that we have used $K=3$ in all experiments shown in this work.
Moreover, $\text{NCC}[f,g]$ denotes the normalized cross-correlation of the two images $f$ and $g$, i.e.
\[\text{NCC}[f,g]:=\frac{1}{\abs{\Omega}}\int_\Omega\frac{\brac{f-\overline{f}}}{\sigma_f}\frac{\brac{g-\overline{g}}}{\sigma_g}\dx,\]
where $\overline{f}=\frac{1}{\abs{\Omega}}\int_\Omega f\dx$ is the mean value of $f$ and $\sigma_f=\sqrt{\frac{1}{\abs{\Omega}}\int_\Omega \brac{f-\overline{f}}^2\dx}$ the standard deviation of $f$. $D\phi$ denotes the Jacobian matrix of $\phi$ and $\identity$ is the identify matrix. Thus, $D\phi-\identity$ is the Jacobian of the displacement component of the deformation $\phi$. As norm for this matrix, we use the Frobenius norm, i.e.\ the Euclidean norm of the matrix interpreted as long vector. $\lambda>0$ is a parameter that controls the strength of the regularization induced by the norm of the Jacobian of the displacement.
The advantage of the alternating minimization strategy is that the deformations $\phi_i$ can be updated one at a time, since the deformations are not directly coupled with each other.
The disadvantage is that the optimization of the deformations is tied to the coordinate system of $f$. This effect can be illustrated with a synthetic input series.

\begin{figure*}[h]
	\setlength{\imgwidth}{0.23\linewidth}
	\setlength{\tabcolsep}{3pt}
	\centering
\begin{tabular}{ccc}
$f_1$ & $f_{10}$ & $f_{19}$\\
\begin{tikzonimage}[width=\imgwidth]{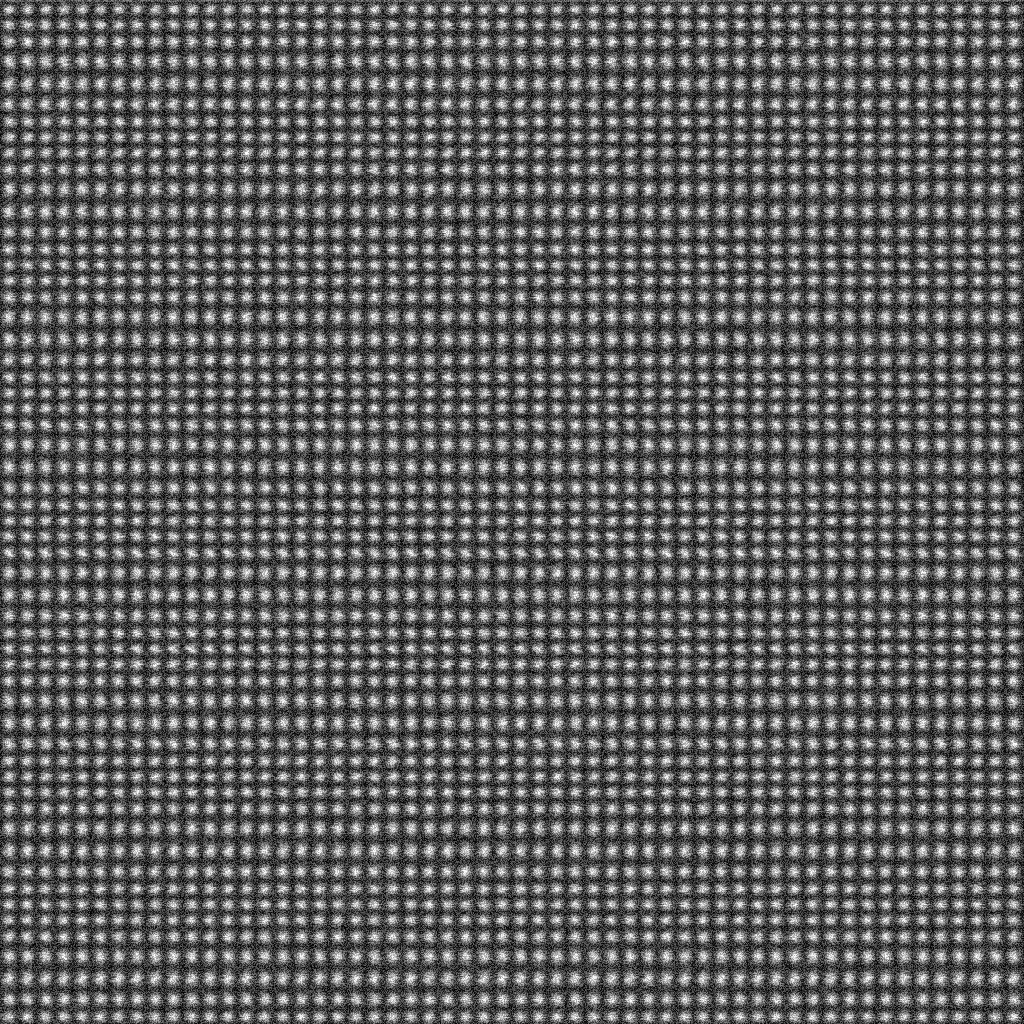}
\node[color=white] at (0.075,0.925) {\Large a)};
\end{tikzonimage}&
\begin{tikzonimage}[width=\imgwidth]{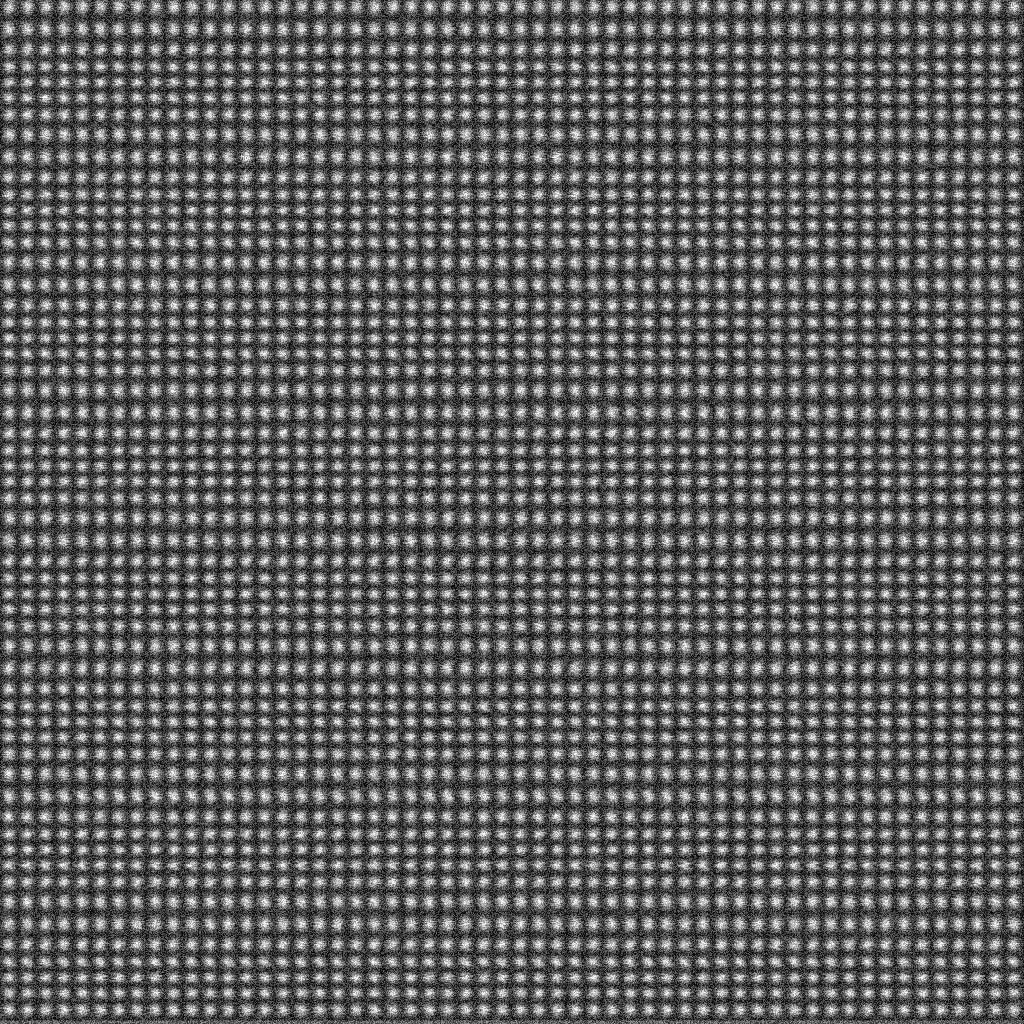}
\node[color=white] at (0.075,0.925) {\Large b)};
\end{tikzonimage}&
\begin{tikzonimage}[width=\imgwidth]{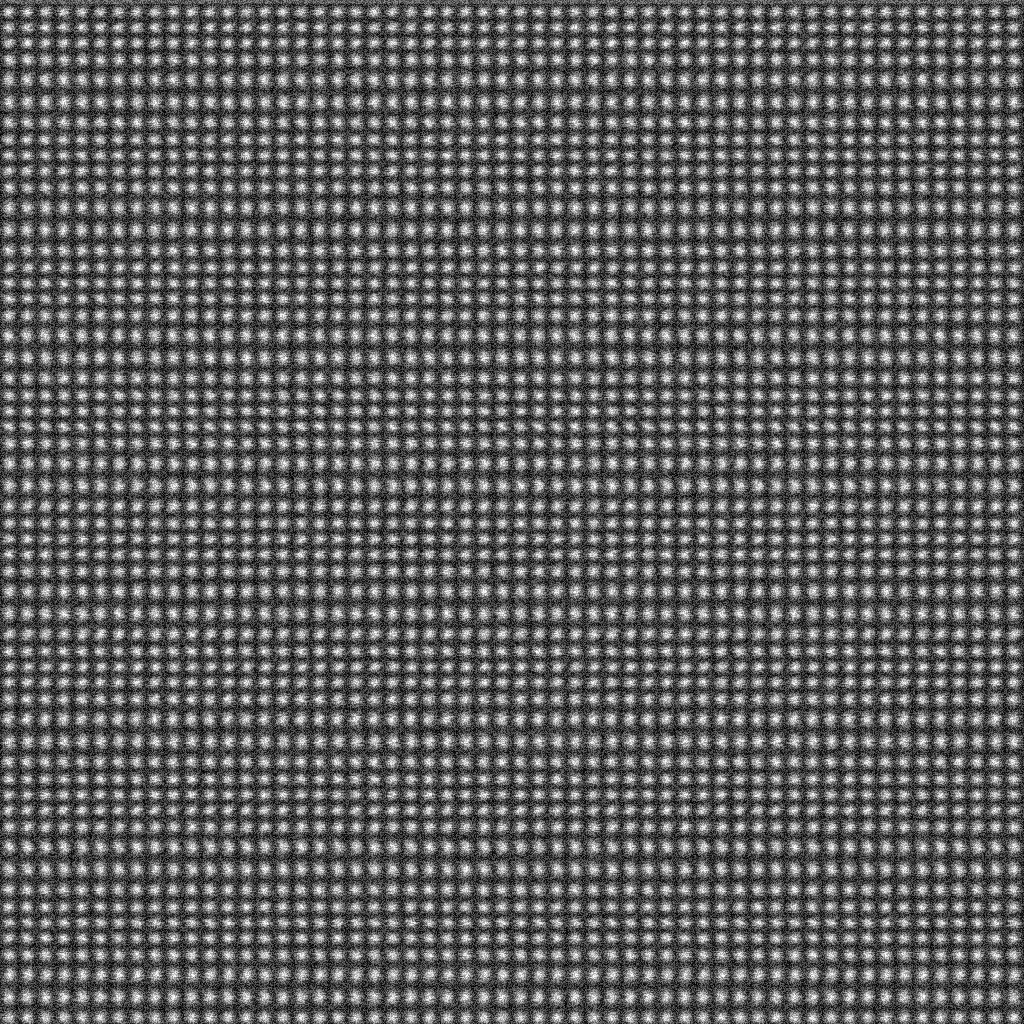}
\node[color=white] at (0.075,0.925) {\Large c)};
\end{tikzonimage}\\
\begin{tikzonimage}[width=\imgwidth]{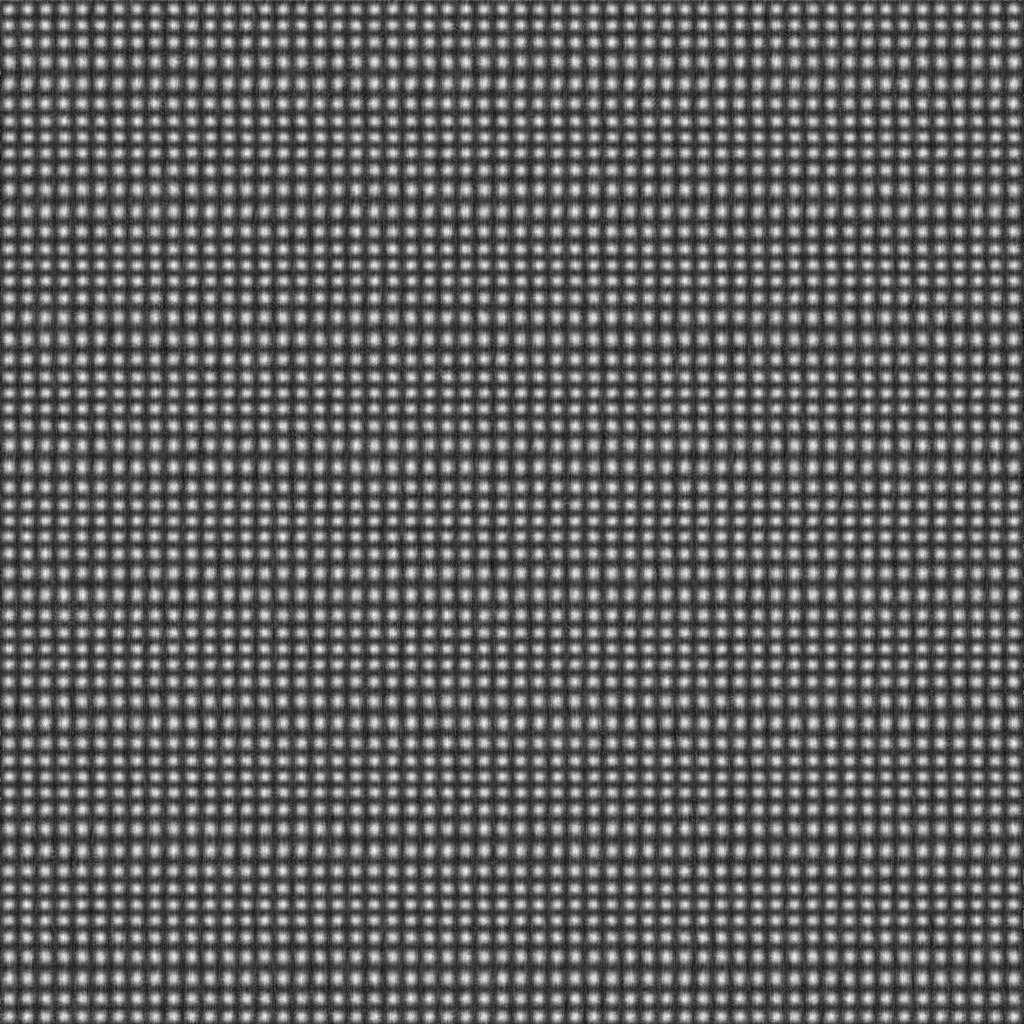}
\node[color=white] at (0.075,0.925) {\Large d)};
\end{tikzonimage}&
\begin{tikzonimage}[width=\imgwidth]{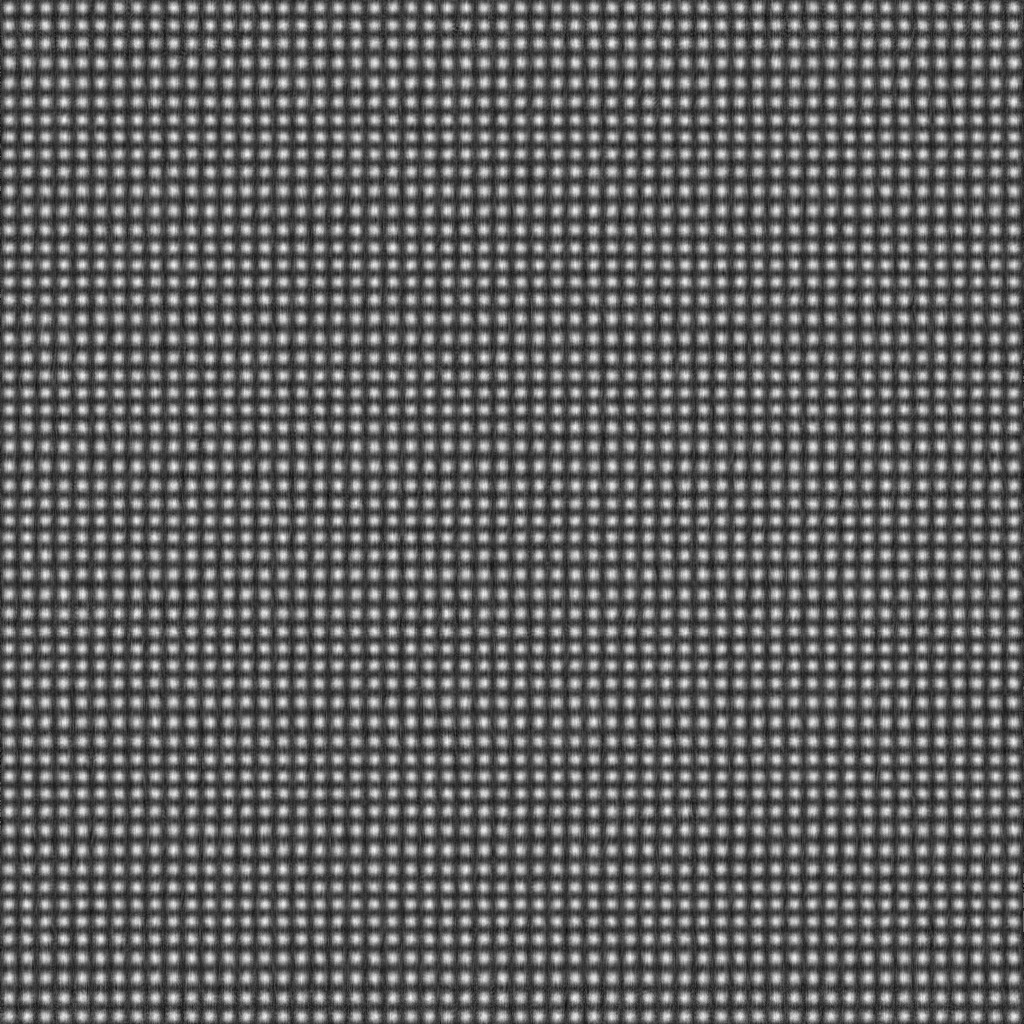}
\node[color=white] at (0.075,0.925) {\Large e)};
\end{tikzonimage}&
\begin{tikzonimage}[width=\imgwidth]{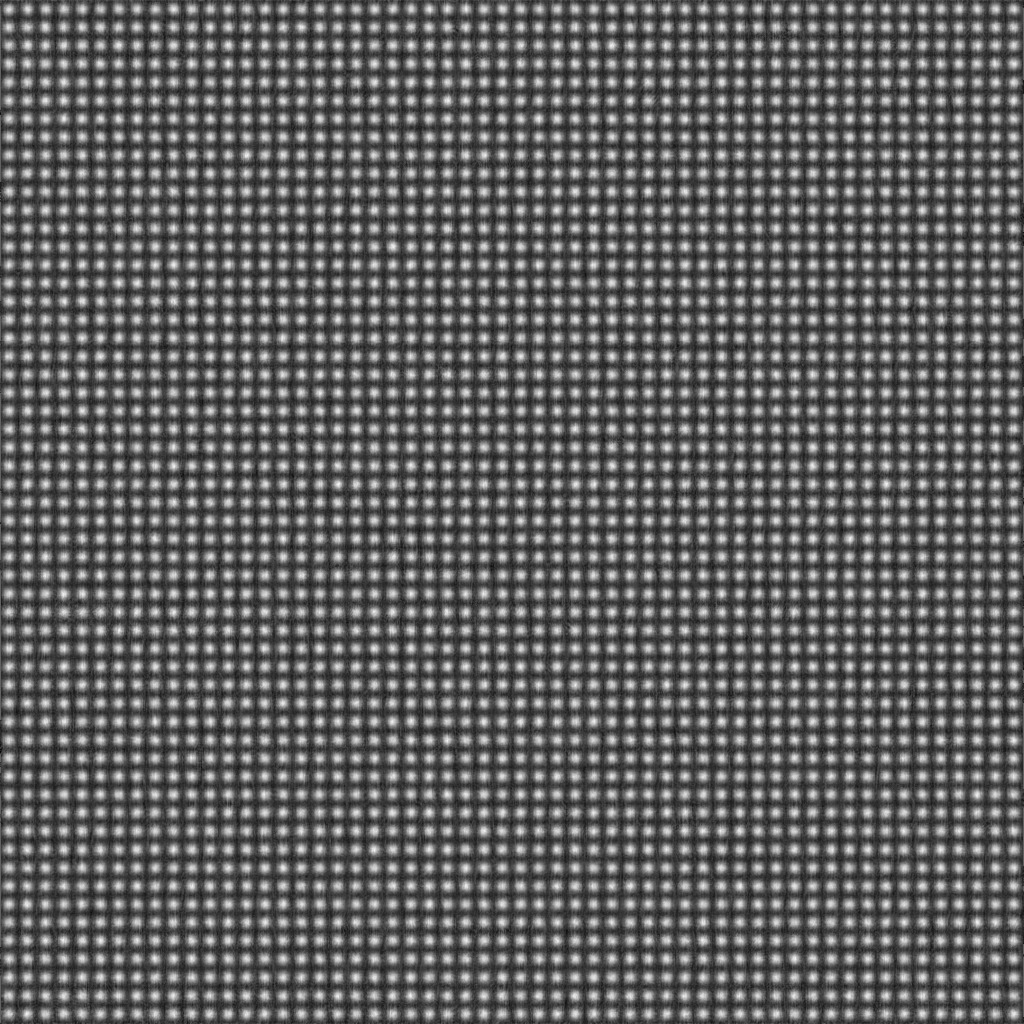}
\node[color=white] at (0.075,0.925) {\Large f)};
\end{tikzonimage}\\
$f$, no bias-correction & $f$, no bias-correction & $f$ with bias-correction\\
$\lambda=10^{-2}$ & $\lambda=20$ & $\lambda=10^{-2}$
\end{tabular}
\caption{Three frames of an artificial 20 frame input image series (a, b and c) and the corresponding reconstruction d+e) without and f) with bias-correction and different parameters. Here, ``no bias-correction'' refers to the method from \cite{BeBiBl13}.}
\label{fig:Arti}
\end{figure*}
We consider a series of 20 synthetic images where each frame contains random fast scan noise in the horizontal direction with a magnitude of $\sim$25$\%$ of the full width at half maximum (FWHM) of the Gaussian-shaped atomic column width. In the vertical direction, each frame is imposed with a sinusoidal wave expanding and compressing the atomic columns, where every four unit cells are expanded, the other four are compressed, as shown in the top row of \cref{fig:Arti}. This simulates strong instabilities resulting from slow scan noise and an imperfect repositioning of the electron beam after the fly back.
As to be expected by a non-rigid averaging approach, \cite{BeBiBl13} noticeably reduces the intensity noise without blurring the atomic columns, cf.\ lower left of \cref{fig:Arti}.
Due to the alternating minimization strategy in \cite{BeBiBl13}, the alignment is biased towards the coordinate system of the first frame of the series though, i.e.\ the average is essentially a denoised version of the first frame inheriting all of its slow scan noise. In contrast, the bias-corrected average obtained with the new approach, which will be discussed below, contains no discernible slow scan noise, but shows an atomic grid that has no distortion visible to the naked eye. It should be mentioned that the overall reduction of distortions with respect to single frame observations largely depends on the number of frames $n$ in an image series and scales with $1/\sqrt{n}$ as discussed in detail by Jones et al.~\cite{JoWeNo2017}.

Note that this series is not a realistic dataset and does not properly reflect the average reconstruction performance of \cite{BeBiBl13}, but was solely constructed to artificially illustrate and emphasize the initialization bias of the alternating strategy. Moreover, the regularization parameter was chosen to amplify the effect. We have used $\lambda=10^{-2}$ and additional regularization with the Laplacian of the deformation scaled by $\lambda_\text{lap}=10^{-4}$ to get sufficiently smooth deformations despite the low value of $\lambda$. When using $\lambda=20$ and $\lambda_\text{lap}=0$, which is what we typically use and also use for all other experiments, the effect of the bias is much lower, cf.\ \cref{fig:Arti}c. Still, the distortion of the grid is visible to the naked eye.
The bias was negligible for the datasets considered in \cite{BeBiBl13,YaBeDa14} as apparent from the precision analysis in \cite{YaBeDa14}, since the individual frames there do not have a strong non-local distortion, which is also the case for many other STEM datasets. This is different for the data considered here, but the effect can be corrected as follows. After determining deformations $\phi_i$ such that the deformed $i$-th frame $f_i\circ\phi_i$ resembles the average $f$ via \cite{BeBiBl13}, these deformations are reduced by computing a deformation $\psi$ that minimizes
\[N[\psi,\phi_1,\ldots,\phi_n]:=\sum_{i=1}^n\int_\Omega\Abs{\phi_i(\psi(x))-x}^2\dx.\]
This way, $\psi$ collects the inverse of the bias, and the bias is removed by replacing $\phi_i$ with $\phi_i\circ\psi$
and $f$ with $f\circ\psi$. After this additional step, the new average is computed as in \cite{BeBiBl13}. Note that the computation of $\psi$ introduces a direct coupling of the $\phi_i$, which \cite{BeBiBl13} is lacking.
\cref{alg:ExtendedAlternatingMinimization} summarizes the resulting minimization strategy. The registration and averaging steps are as in \cite{BeBiBl13}, the bias-reduction step is the addition proposed here. As in \cite{BeBiBl13}, we use $0.1\lambda$ instead of $\lambda$ as regularization parameter for $k\geq2$.
When applied to the artificial example we considered above (\cref{fig:Arti}), the reduction removes the bias towards the first frame and leads to the expected distortion-free reconstruction.
\begin{algorithm}
\caption{Extended alternating minimization strategy}
\label{alg:ExtendedAlternatingMinimization}
\begin{algorithmic}
  \STATE{Initialize the average with $f:=f_1$.}
  \FOR{$k=1,\ldots,K$}
    \STATE{\% \textit{Registration step}}
    \STATE{For $i=1,\ldots,n$, compute $\phi_i$ such that $f_i\circ \phi_i\approx f$.}
    \STATE{}
    \STATE{\% \textit{Bias-reduction step}}
    \STATE{Compute $\psi$ that minimizes $N[\psi,\phi_1,\ldots,\phi_n]$.}
    \STATE{For $i=1,\ldots,n$, replace $\phi_i$ with $\phi_i\circ\psi$.}
    \STATE{}
    \STATE{\% \textit{Averaging step}}
    \STATE{Update the average $f$ based on fixed $(\phi_1,\ldots,\phi_n)$.}
  \ENDFOR
\end{algorithmic}
\end{algorithm}

Structurally, the extended algorithm not only introduces a direct coupling of the $\phi_i$, but can be interpreted as way to solve a two-level optimization problem: The minimizers of $E[f,\phi_1,\ldots,\phi_n]$ are not unique, i.e.\ if $(f,\phi_1,\ldots,\phi_n)$ minimizes $E$ and $\psi$ is a suitable coordinate transform, e.g.\ a translation, then $(f\circ\psi,\phi_1\circ\psi,\ldots,\phi_n\circ\psi)$ also minimizes $E$.
From these possible minimizers, the algorithm from \cite{BeBiBl13} selects that minimizer where $\phi_1\approx\identityMapping$, i.e.\ it inherits the coordinate system of $f_1$. Here, $\identityMapping$ denotes the identity mapping.
The extended algorithm selects the minimizer where $N[\identityMapping,\phi_1,\ldots,\phi_n]$ is minimal.
This is a consequence of $N[\psi,\phi_1,\ldots,\phi_n]=N[\identityMapping,\phi_1\circ\psi,\ldots,\phi_n\circ\psi]$ and that $E[f,\phi_1,\ldots,\phi_n]\approx E[f\circ\psi,\phi_1\circ\psi,\ldots,\phi_n\circ\psi]$ for sufficiently smooth $\psi$.

\section{Results}
\subsection{Lattice strain determination at coherent interfaces}
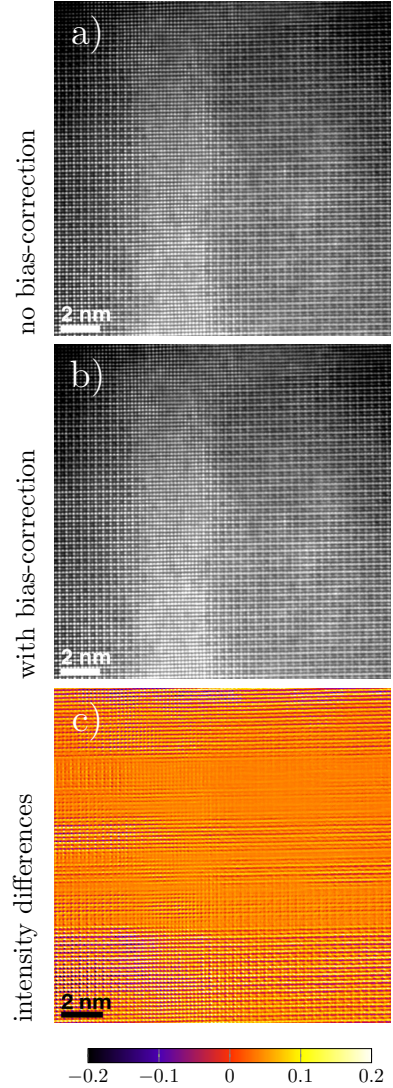
\begin{figure}[ht]
	\setlength{\imgwidth}{0.5\linewidth}
	\centering
	\setlength{\tabcolsep}{0.75ex}
  \begin{tabular}{cc}
	\rotatebox{90}{no bias-correction}&
	\begin{tikzonimage}[width=\imgwidth]{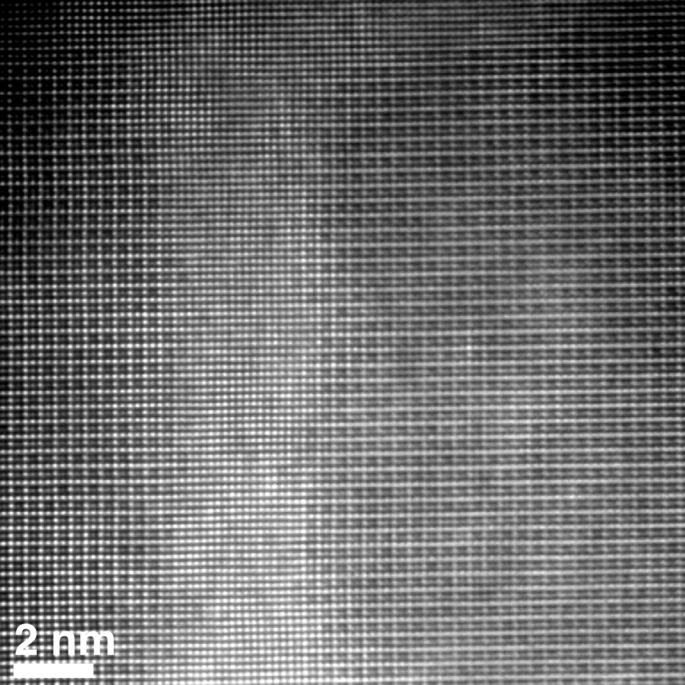}
	\node[color=white] at (0.1,0.9) {\Large a)};
	\end{tikzonimage}\\
	\rotatebox{90}{with bias-correction}&
	\begin{tikzonimage}[width=\imgwidth]{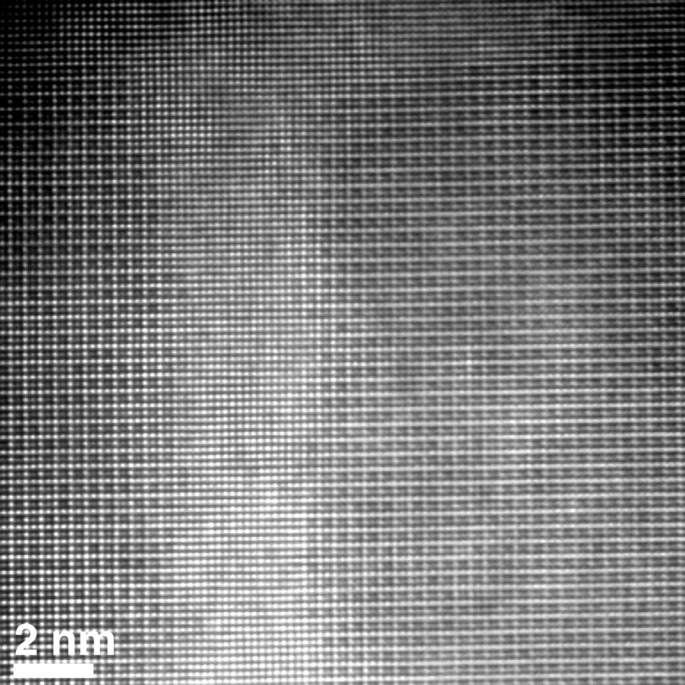}
	\node[color=white] at (0.1,0.9) {\Large b)};
	\end{tikzonimage}\\
	\rotatebox{90}{intensity differences}&
	\begin{tikzonimage}[width=\imgwidth]{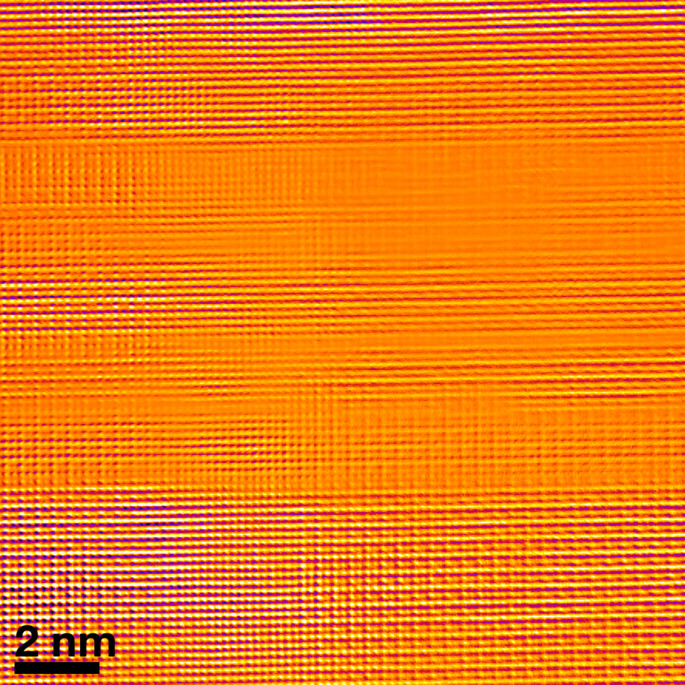}
	\node[color=white] at (0.1,0.9) {\Large c)};
	\end{tikzonimage}\\
		&
		\resizebox{\imgwidth}{!}{
		\begin{tikzpicture}
	\begin{axis}[
	    hide axis,
	    scale only axis,
	    height=0pt,
	    width=0pt,
			xmin=0,xmax=1,
			ymin=0,ymax=1,
	    colormap/thermal,
	    colorbar horizontal,
	    point meta min=-0.2,
	    point meta max=0.2,
	    colorbar style={
	        width=5cm,
					height=0.25cm,
					xtick distance=0.1
	    }]
	    \addplot [draw=none] coordinates {(0,0)};
	\end{axis}
	\end{tikzpicture}}
	\end{tabular}
	\caption{\footnotesize Average of raw images from a serial acquisition scheme of 20 individual images  aligned by non-rigid registration without a) bias-correction and b) with bias-correction. The normalized difference image of a) and b) is given in c).}
	\label{fig:kappa}
\end{figure}

The novel extension of the minimization strategy is tested on two experimental datasets. In the first example, the method is applied to coherent interfaces between two $\kappa$-carbide nano-precipitates embedded in an face centered cubic (fcc) Fe matrix. The comparison of the non-rigid registration without and with bias-correction of a series of 20 individual images recorded with a dwell time of 2 $\mu$s per pixel is illustrated in \cref{fig:kappa}. The ordered nature of the $\kappa$-carbide is clearly seen on the left and right side of the images in Figs.\ \ref{fig:kappa} a) and b) due to the Z-contrast in the HAADF images.

In the viewing direction of [001], the superlattice order of the Fe$_{3}$AlC $\kappa$-phase is represented by horizontal planes composed of Fe-rich (Z=26) columns and planes with alternating Fe and Al (Z=13) columns. The atomic columns in the vertical fcc-Fe matrix channel exhibit a homogeneous intensity distribution. In previous work, it was shown that the lattice parameter mismatch between the precipitate phases and the matrix leads to a tetragonal distortion of the narrow fcc-Fe matrix channel, since it is coherently constrained \cite{LiYaDe18}.

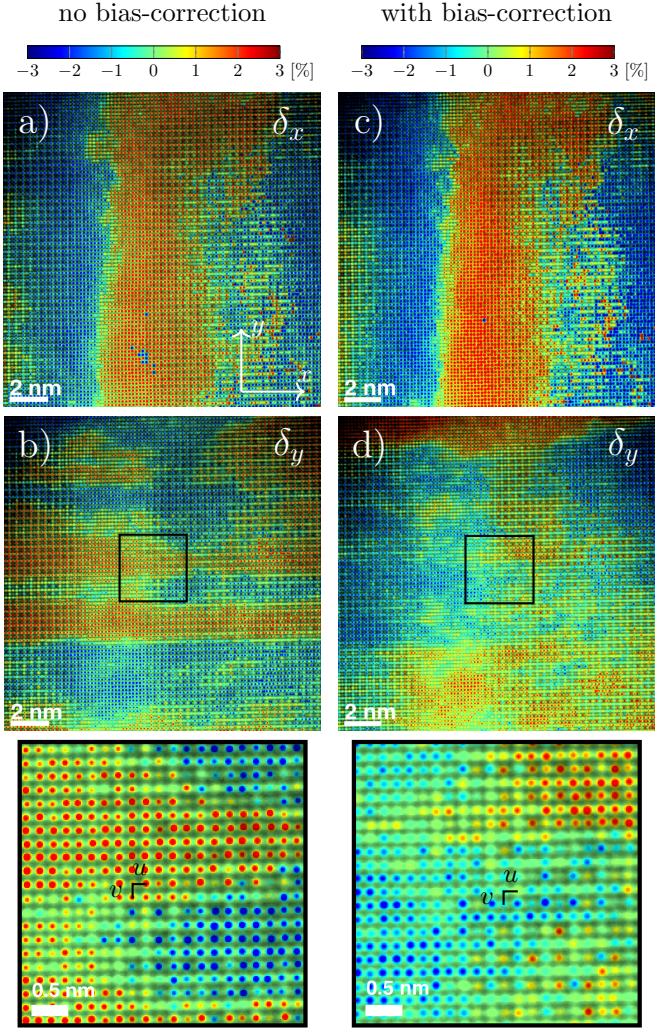
\begin{figure}[ht]
	\setlength{\imgwidth}{0.47\linewidth}
	\centering
	\setlength{\tabcolsep}{0.75ex}
  \begin{tabular}{cc}
	no bias-correction&
	with bias-correction\\
	\resizebox{\imgwidth}{!}{
	\begin{tikzpicture}
\begin{axis}[
    hide axis,
    scale only axis,
    height=0pt,
    width=0pt,
		xmin=0,xmax=1,
		ymin=0,ymax=1,
    colormap/jet,
    colorbar horizontal,
    point meta min=-3,
    point meta max=3,
    colorbar style={
        width=5cm,
				height=0.25cm,
        xtick={-3,-2,-1,0,1,2,3}
    }]
    \addplot [draw=none] coordinates {(0,0)};
\end{axis}
\end{tikzpicture}\raisebox{1ex}{[\%]}
}&
\resizebox{\imgwidth}{!}{
\begin{tikzpicture}
\begin{axis}[
	hide axis,
	scale only axis,
	height=0pt,
	width=0pt,
	xmin=0,xmax=1,
	ymin=0,ymax=1,
	colormap/jet,
	colorbar horizontal,
	point meta min=-3,
	point meta max=3,
	colorbar style={
			width=5cm,
			height=0.25cm,
			xtick={-3,-2,-1,0,1,2,3}
	}]
	\addplot [draw=none] coordinates {(0,0)};
\end{axis}
\end{tikzpicture}\raisebox{1ex}{[\%]}
}\\
	\begin{tikzonimage}[width=\imgwidth]{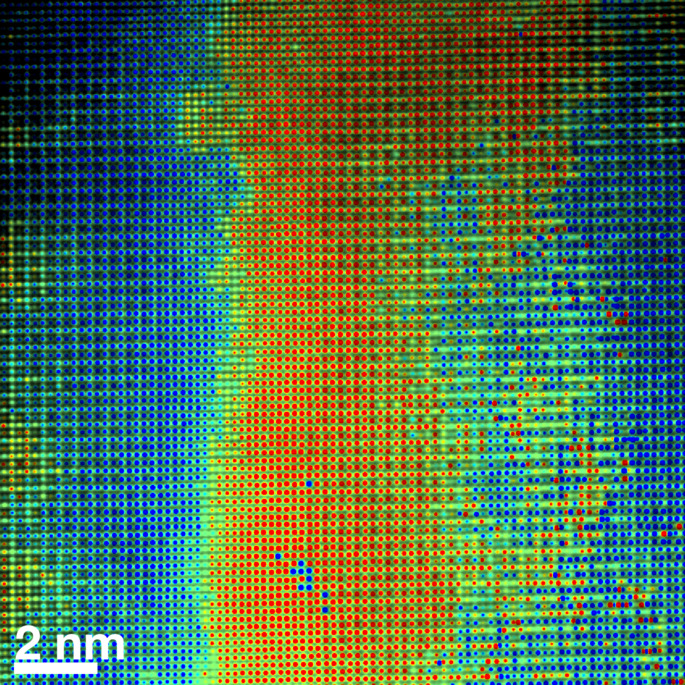}
	\node[color=white] at (0.1,0.9) {\Large a)};
	\node[color=white] at (0.9,0.9) {\Large $\delta_x$};
	\draw[->,color=white,thick] (0.75, 0.05) -- (0.95, 0.05);
	\node[above,color=white] at (0.95, 0.05) {$x$};
	\draw[->,color=white,thick] (0.75, 0.05) -- (0.75, 0.25);
	\node[right,color=white] at (0.75, 0.25) {$y$};
	\end{tikzonimage}&
	\begin{tikzonimage}[width=\imgwidth]{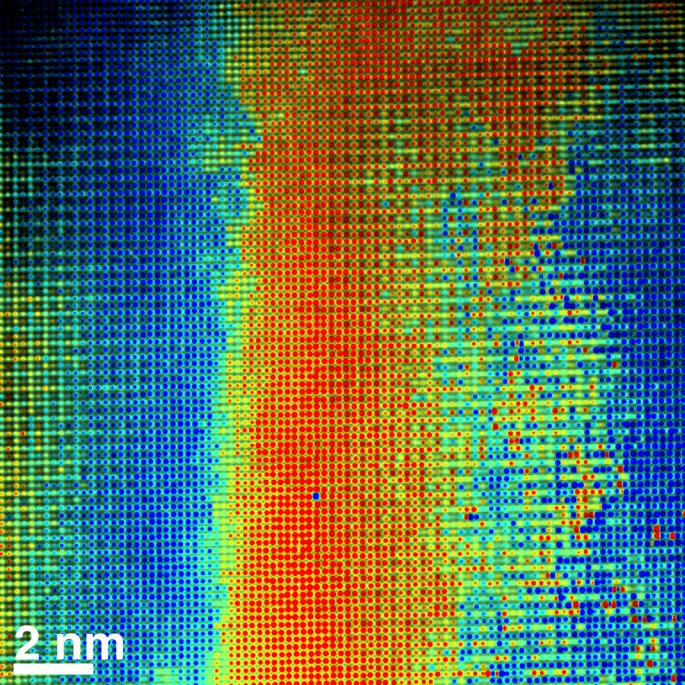}
	\node[color=white] at (0.1,0.9) {\Large c)};
	\node[color=white] at (0.9,0.9) {\Large $\delta_x$};
	\end{tikzonimage}\\
	\begin{tikzonimage}[width=\imgwidth]{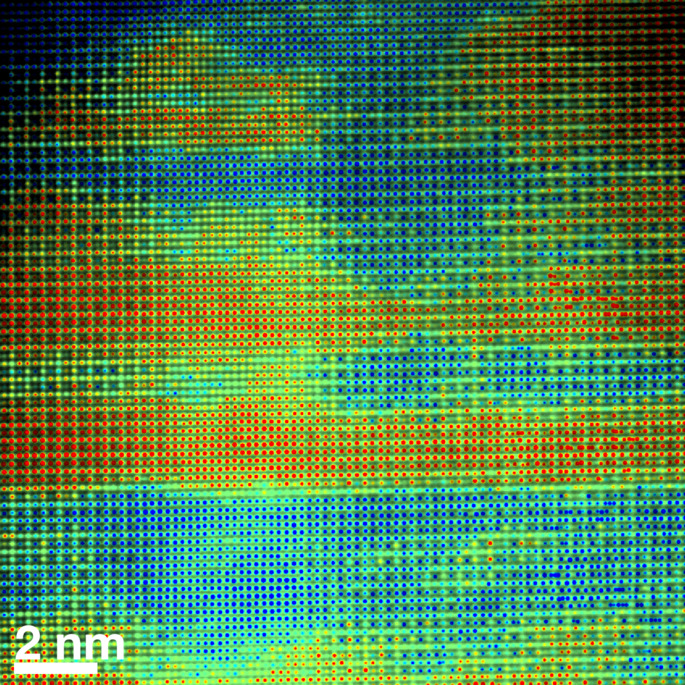}
	\node[color=white] at (0.1,0.9) {\Large b)};
	\node[color=white] at (0.9,0.9) {\Large $\delta_y$};
	\draw[thick] (0.364566,1-0.377327) rectangle (0.575809,1-0.588832);
	\end{tikzonimage}&
	\begin{tikzonimage}[width=\imgwidth]{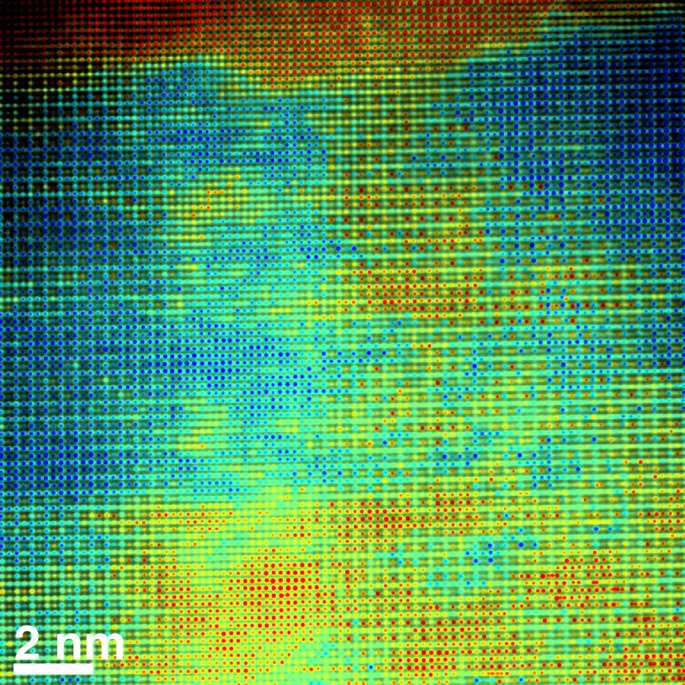}
	\node[color=white] at (0.1,0.9) {\Large d)};
	\node[color=white] at (0.9,0.9) {\Large $\delta_y$};
	\draw[thick] (0.401361,1-0.381136) rectangle (0.615646,1-0.594915);
	\end{tikzonimage}\\
	\begin{tikzonimage}[width=0.9\imgwidth]{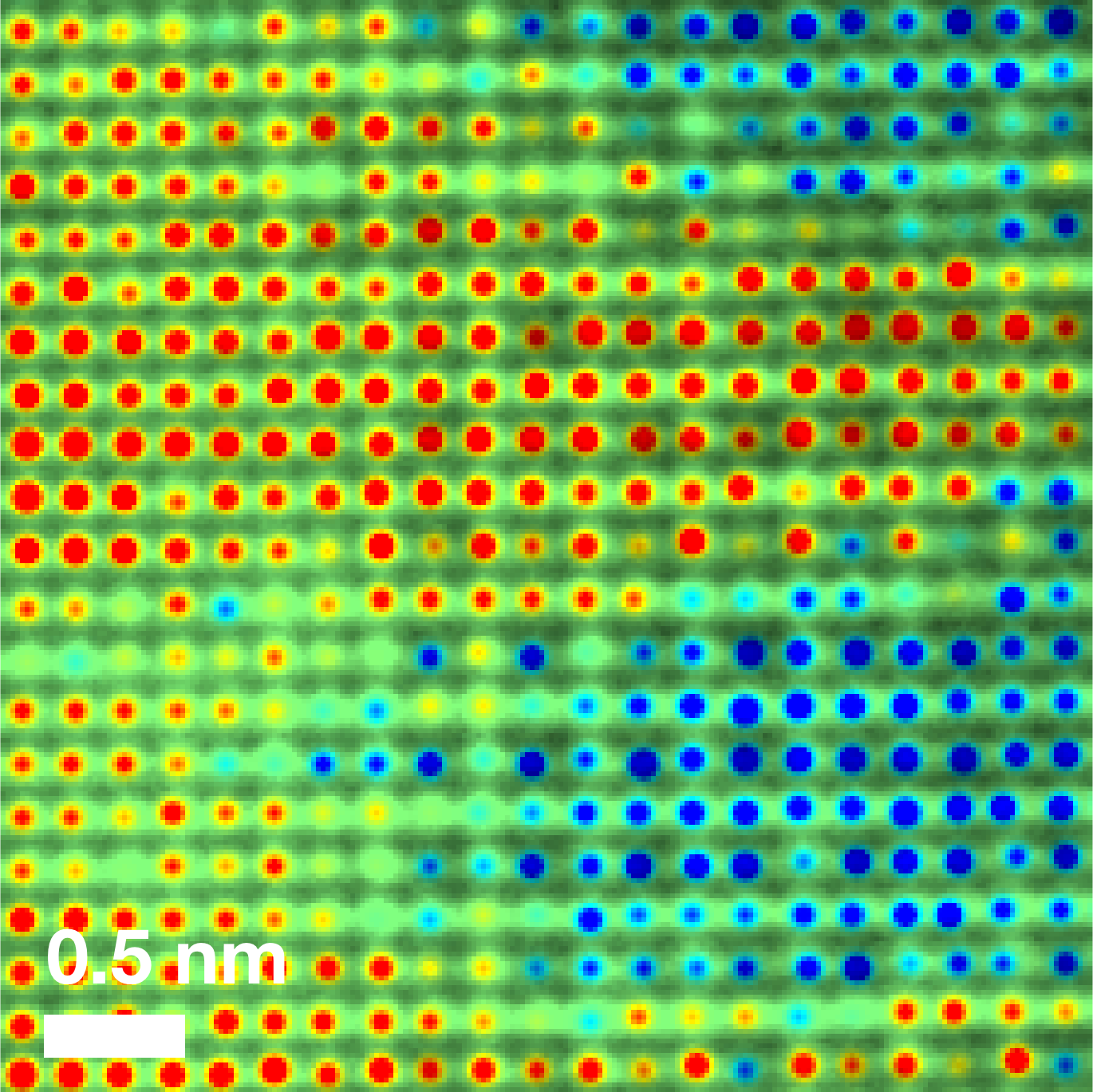}
	\tikzmath{
		\numX = 1370;
		\numY = 1369;
		\x1 = 541/(\numX-1);
		\y1 = 1-686/(\numY-1);
		\x2 = 609/(\numX-1);
		\y2 = \y1;
		\x3 = \x1;
		\y3 = 1-756/(\numY-1);
	}
	\draw[thick] (\x3, \y3) -- (\x1, \y1) -- (\x2, \y2);
	\node[above] at (0.5*\x1+0.5*\x2, 0.5*\y1+0.5*\y2) {$u$};
	\node[left] at (0.5*\x1+0.5*\x3, 0.5*\y1+0.5*\y3) {$v$};
	\draw[ultra thick] (0,0) rectangle (1,1);
	\end{tikzonimage}
	&
	\begin{tikzonimage}[width=0.9\imgwidth]{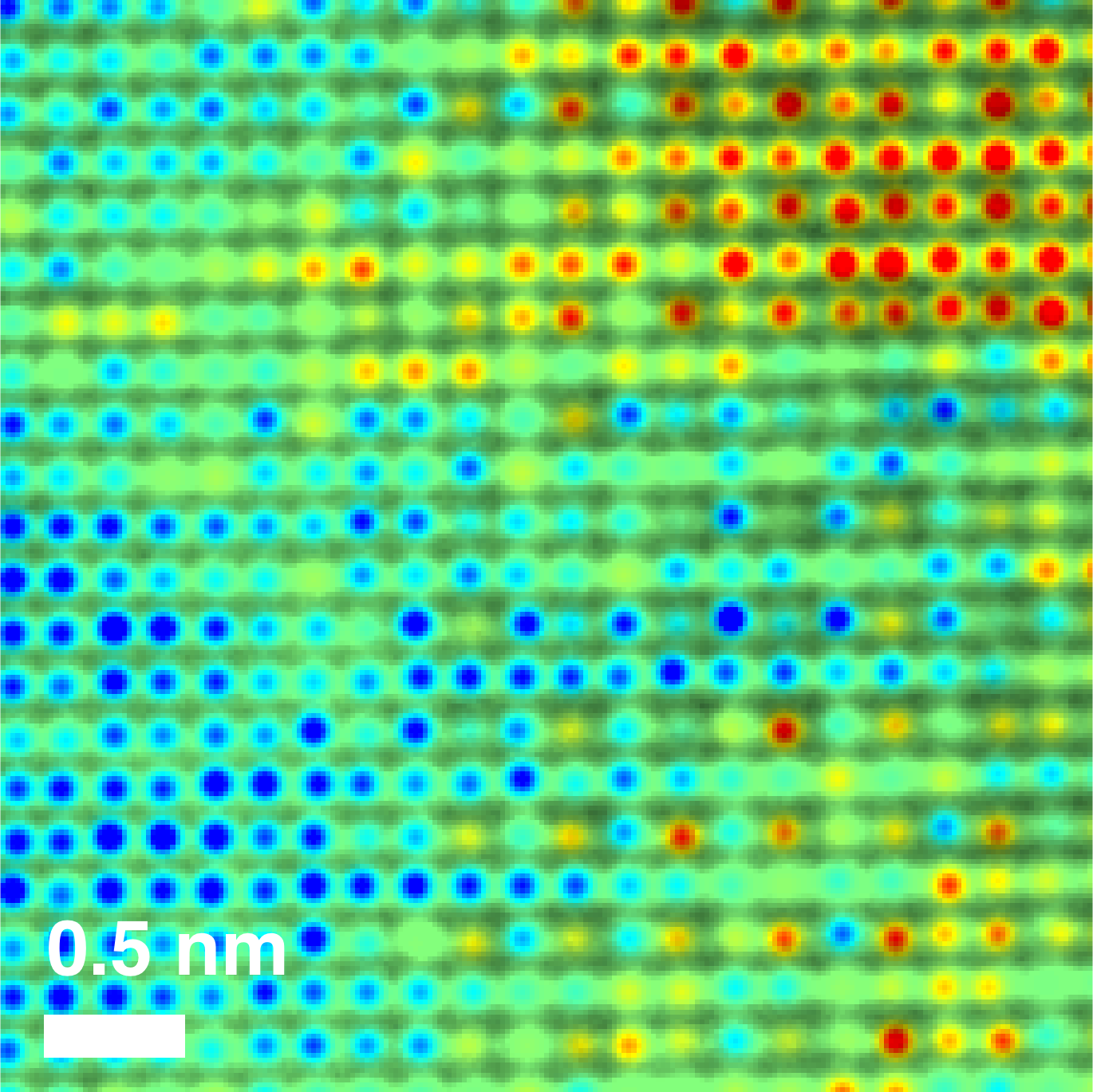}
	\tikzmath{
		\numX = 1370;
		\numY = 1369;
		\x1 = 717/(\numX-1);
		\y1 = 1-721/(\numY-1);
		\x2 = 785/(\numX-1);
		\y2 = \y1;
		\x3 = \x1;
		\y3 = 1-786/(\numY-1);
	}
	\draw[thick] (\x3, \y3) -- (\x1, \y1) -- (\x2, \y2);
	\node[above] at (0.5*\x1+0.5*\x2, 0.5*\y1+0.5*\y2) {$u$};
	\node[left] at (0.5*\x1+0.5*\x3, 0.5*\y1+0.5*\y3) {$v$};
	\draw[ultra thick] (0,0) rectangle (1,1);
	\end{tikzonimage}
	\end{tabular}
	\caption{\footnotesize Displacement maps determined from fitted atomic column peak positions. The left column shows the dataset for the approach without bias-correction, the right column for the approach with bias-correction. The $x$-components $\delta_{x}$ are given in a) and c), the $y$-components $\delta_{y}$ in b) and d), respectively. The bottom row of images are the magnified central regions of the images labeling the local lattice vectors $u$ and $v$.}
	\label{fig:kappaDisp}
\end{figure}

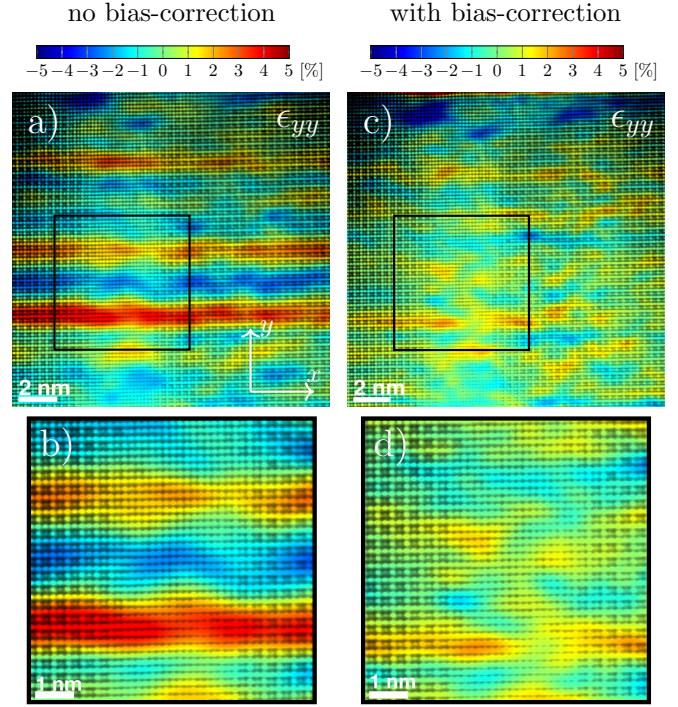
\begin{figure}[ht]
	\setlength{\imgwidth}{0.47\linewidth}
	\centering
	\setlength{\tabcolsep}{0.75ex}
  \begin{tabular}{cc}
	no bias-correction&
	with bias-correction\\
		\resizebox{\imgwidth}{!}{
		\begin{tikzpicture}
	\begin{axis}[
	    hide axis,
	    scale only axis,
	    height=0pt,
	    width=0pt,
			xmin=0,xmax=1,
			ymin=0,ymax=1,
	    colormap/jet,
	    colorbar horizontal,
	    point meta min=-5,
	    point meta max=5,
	    colorbar style={
	        width=5cm,
					height=0.25cm,
	        xtick={-5,-4,-3,-2,-1,0,1,2,3,4,5}
	    }]
	    \addplot [draw=none] coordinates {(0,0)};
	\end{axis}
	\end{tikzpicture}\raisebox{1ex}{[\%]}
	}&
	\resizebox{\imgwidth}{!}{
	\begin{tikzpicture}
\begin{axis}[
		hide axis,
		scale only axis,
		height=0pt,
		width=0pt,
		xmin=0,xmax=1,
		ymin=0,ymax=1,
		colormap/jet,
		colorbar horizontal,
		point meta min=-5,
		point meta max=5,
		colorbar style={
				width=5cm,
				height=0.25cm,
				xtick={-5,-4,-3,-2,-1,0,1,2,3,4,5}
		}]
		\addplot [draw=none] coordinates {(0,0)};
	\end{axis}
	\end{tikzpicture}\raisebox{1ex}{[\%]}
	}\\
	\begin{tikzonimage}[width=\imgwidth]{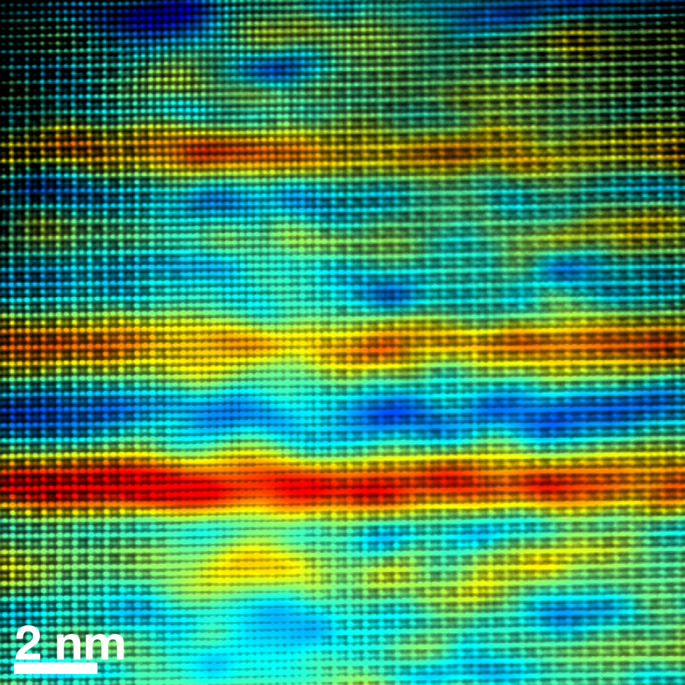}
	\node[color=white] at (0.1,0.9) {\Large a)};
	\node[color=white] at (0.9,0.9) {\Large $\epsilon_{yy}$};
	\draw[->,color=white,thick] (0.75, 0.05) -- (0.95, 0.05);
	\node[above,color=white] at (0.95, 0.05) {$x$};
	\draw[->,color=white,thick] (0.75, 0.05) -- (0.75, 0.25);
	\node[right,color=white] at (0.75, 0.25) {$y$};
	\draw[thick] (0.132231,1-0.391185) rectangle (0.557851,1-0.816804);
	\end{tikzonimage}&
	\begin{tikzonimage}[width=\imgwidth]{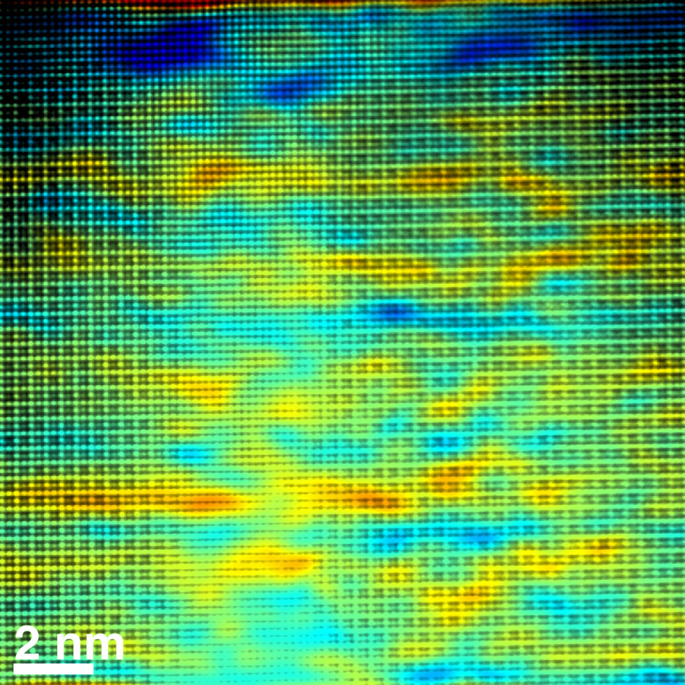}
	\node[color=white] at (0.1,0.9) {\Large c)};
	\node[color=white] at (0.9,0.9) {\Large $\epsilon_{yy}$};
	\draw[thick] (0.147586,1-0.391724) rectangle (0.572414,1-0.817931);
	\end{tikzonimage}\\
	\begin{tikzonimage}[width=.9\imgwidth]{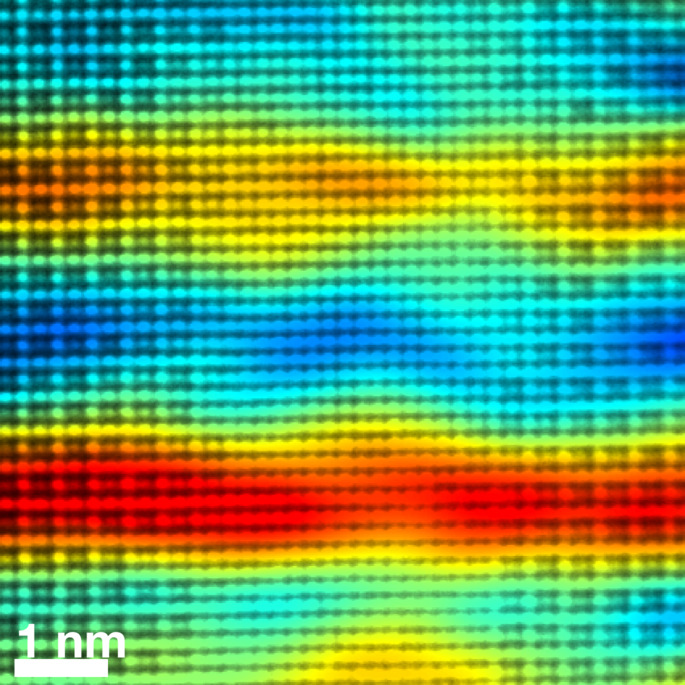}
	\node[color=white] at (0.1,0.9) {\Large b)};
	\draw[ultra thick] (0,0) rectangle (1,1);
	\end{tikzonimage}&
	\begin{tikzonimage}[width=.9\imgwidth]{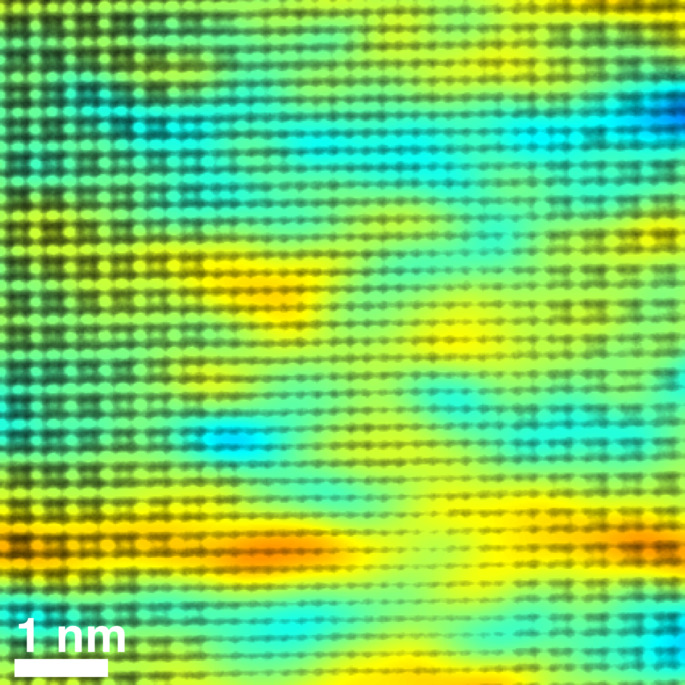}
	\node[color=white] at (0.1,0.9) {\Large d)};
	\draw[ultra thick] (0,0) rectangle (1,1);
	\end{tikzonimage}
	\end{tabular}
	\caption{\footnotesize The left and right column display the $\epsilon_{yy}$ strain component of the non-rigid registration approach without and with bias-correction, respectively. a) and c) illustrate the complete strain maps with the local coordinate system and b) and d) show magnified subregions as indicated by the black rectangles in a) and c).}
	\label{fig:kappaStrain}
\end{figure}

The difference image of the two non-rigidly registered datasets (\cref{fig:kappa} c)) reveals subtle differences especially in horizontal stripes across the entire image. This difference in the non-rigid registration approaches without and with bias-correction results from slow scan noise, since the slow scan direction aligns with the horizontal direction of the image coordinate system. The impact on atomic column displacements is further investigated by first fitting the peak positions with non-linear Gaussian functions. The peak maximum is taken as the center position of the atomic columns. The peak displacements are then obtained by determining the deviation of the actual peak positions to that of the mean lattice of the entire image. The peak displacement maps of both non-rigid registration approaches are shown in \cref{fig:kappaDisp} for comparison. The displacement field in the dataset registered without bias-correction is noisier and the $\delta_{y}$ component of \cref{fig:kappaDisp} b) shows pronounced lattice dilation and expansion along the slow scan direction that is largely suppressed in the approach with bias correction shown in \cref{fig:kappaDisp} d). The magnified central regions of the $\delta_{y}$ components also indicate that the bias-correction is a way to effectively reduce the strong non-local distortions. The gradient of the displacement field directly obtains the strain in atomic resolution images \cite{Hytch2008} and it was shown that the lattice of the narrow matrix channel is compressed in the $x$-direction with respect to that of the $\kappa$-precipitates \cite{LiYaDe18}. The strain field analysis will not be repeated here, but the implications especially on slow scan distortions of the obtained $y$-strain components are compared in \cref{fig:kappaStrain}.

\begin{figure}[ht]
\centering
\setlength{\tabcolsep}{1pt}
\begin{tabular}{cc}
	\rotatebox{90}{no bias-correction}&
	\includegraphics{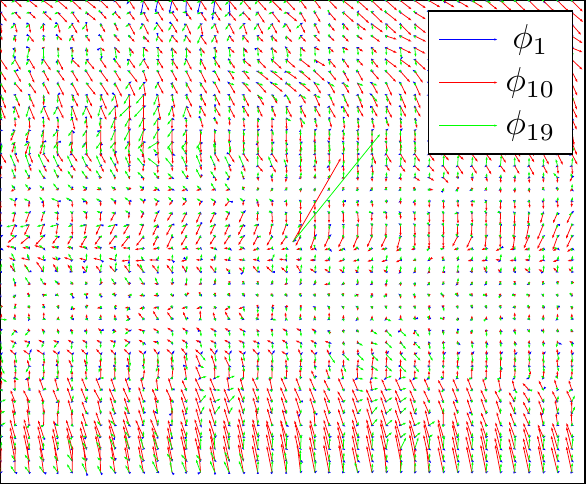}\\
	\rotatebox{90}{with bias-correction}&
	\includegraphics{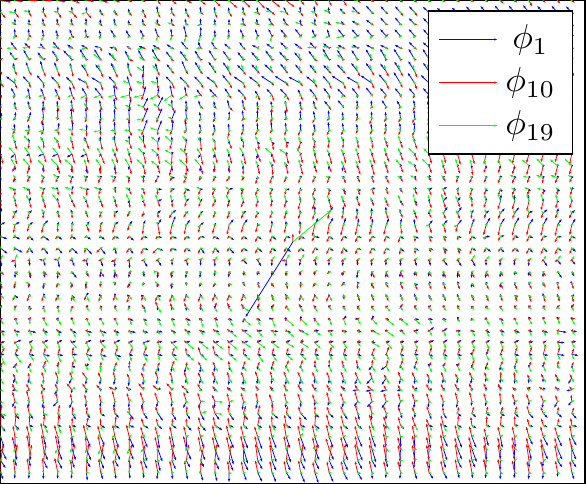}\\
\end{tabular}
\caption{\footnotesize Displacement components of $\phi_1$, $\phi_{10}$ and $\phi_{19}$ corresponding to the reconstruction results from \cref{fig:kappa}. The large arrows in the center show the global translation component of the displacements, the small vectors visualize the local displacement after subtracting the global translation component. All vectors are scaled by a factor of 10 for the purpose of visualization.}
\label{fig:figKappVecPlots}
\end{figure}

The strong non-local distortions in the dataset without bias correction translate into alternating stripes of compressive and dilational strain as seen in Figs.\ \ref{fig:kappaStrain} a) and b). For the current example, these slow scan distortions largely obstruct the interpretation of the physical long range elastic fields in the sample that are crucial to unravel the underlying tetragonal distortion of such narrow matrix channels \cite{LiYaDe18}. The proposed bias-correction is able to reduce these distortions by more than a factor of 2, as illustrated in Figs.\ \ref{fig:kappaStrain} c) and d). It should be mentioned that scan instabilities and in particular slow scan noise are largely depending on the type of microscope and the room conditions on the day of the measurement.
\cref{fig:figKappVecPlots} illustrates the deformations determined without and with bias-correction. Here, the vector in the middle of the plot is the global translation component, i.e.\ the mean displacement component of the deformations $\overline{\phi_i(x)-x}$ (reusing the notation from the definition of $\text{NCC}$), and the others are the local displacements after subtracting the average, i.e.\ $\phi_i(x)-x - \overline{\phi_i(x)-x}$. Without bias-correction, the displacements of the first frame are much smaller than that of frames 10 and 19. With bias-correction, the magnitude of the local displacements is far less frame dependent and more uniformly spread over the frames. Moreover, the average of the global translations is much closer to zero for the case with bias-correction than it is without.

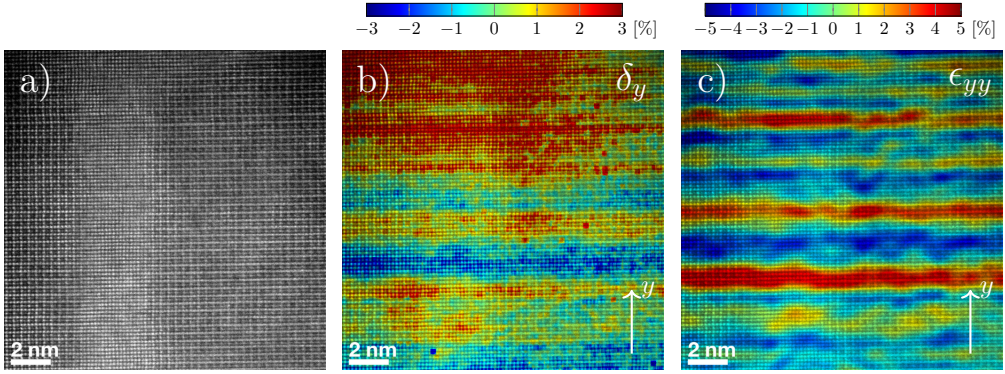
\begin{figure*}[ht]
	\setlength{\imgwidth}{0.23\linewidth}
	\centering
	\setlength{\tabcolsep}{0.75ex}
	\begin{tabular}{ccc}
		\resizebox{\imgwidth}{!}{
			\begin{tikzpicture}
			\end{tikzpicture}}&
		\resizebox{\imgwidth}{!}{
			\begin{tikzpicture}
			\begin{axis}[
			hide axis,
			scale only axis,
			height=0pt,
			width=0pt,
			xmin=0,xmax=1,
			ymin=0,ymax=1,
			colormap/jet,
			colorbar horizontal,
			point meta min=-3,
			point meta max=3,
			colorbar style={
				width=5cm,
				height=0.25cm,
				xtick distance=1,
				scaled ticks=false,
				xticklabel style={/pgf/number format/fixed}
			}]
			\addplot [draw=none] coordinates {(0,0)};
			\end{axis}
			\end{tikzpicture}\raisebox{1ex}{[\%]}
		}&
		\resizebox{\imgwidth}{!}{
			\begin{tikzpicture}
			\begin{axis}[
			hide axis,
			scale only axis,
			height=0pt,
			width=0pt,
			xmin=0,xmax=1,
			ymin=0,ymax=1,
			colormap/jet,
			colorbar horizontal,
			point meta min=-5,
			point meta max=5,
			colorbar style={
				width=5cm,
				height=0.25cm,
				xtick distance=1,
				scaled ticks=false,
				xticklabel style={/pgf/number format/fixed}
			}]
			\addplot [draw=none] coordinates {(0,0)};
			\end{axis}
			\end{tikzpicture}\raisebox{1ex}{[\%]}
		}\\
		\begin{tikzonimage}[width=\imgwidth]{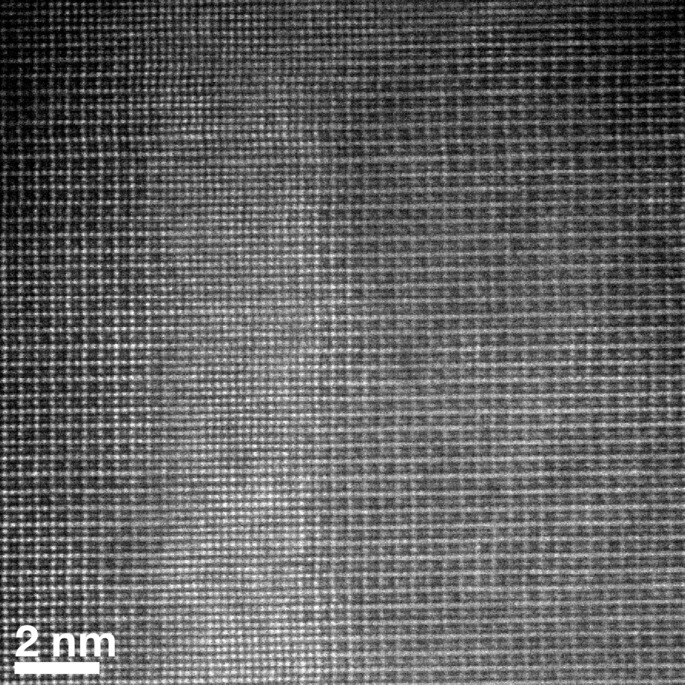}
			\node[color=white] at (0.1,0.9) {\Large a)};
		\end{tikzonimage}&
		\begin{tikzonimage}[width=\imgwidth]{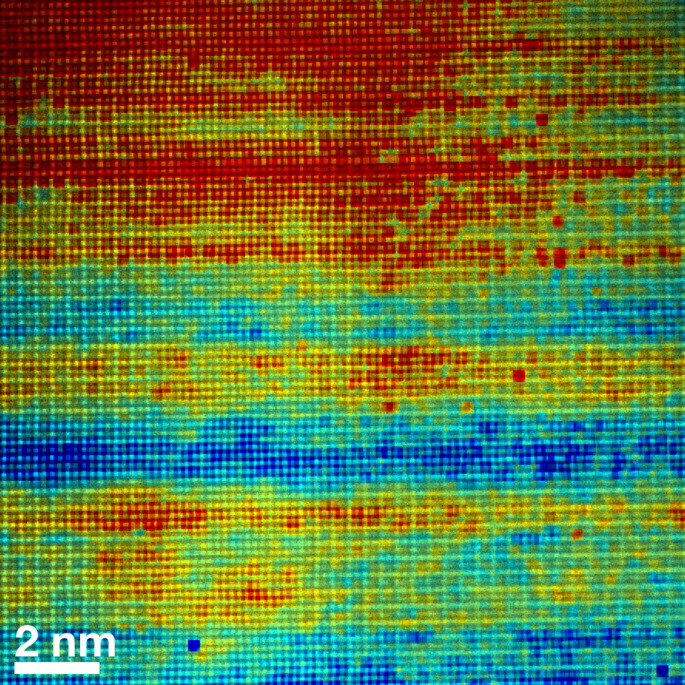}
			\node[color=white] at (0.1,0.9) {\Large b)};
			\node[color=white] at (0.9,0.9) {\Large $\delta_{y}$};
			\draw[->,color=white,thick] (0.9, 0.05) -- (0.9, 0.25);
			\node[right,color=white] at (0.9, 0.25) {$y$};
		\end{tikzonimage}&
		\begin{tikzonimage}[width=\imgwidth]{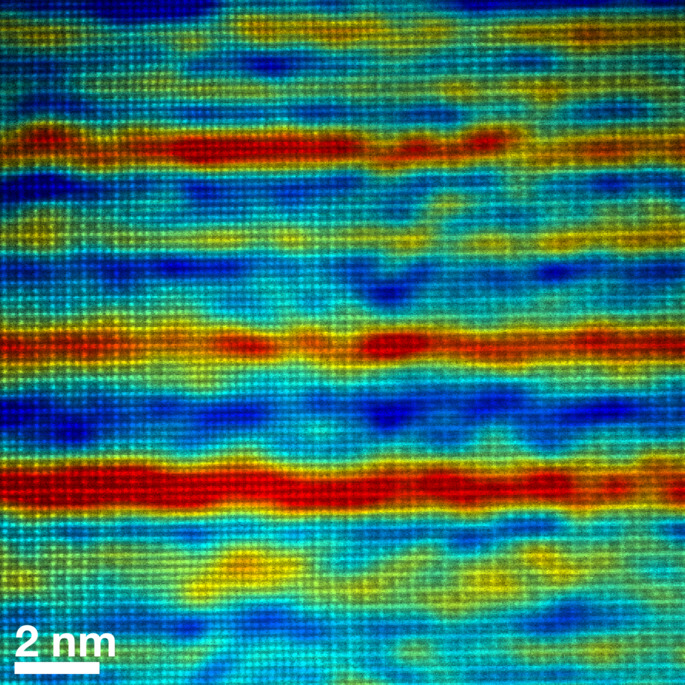}
			\node[color=white] at (0.1,0.9) {\Large c)};
			\node[color=white] at (0.9,0.9) {\Large $\epsilon_{yy}$};
			\draw[->,color=white,thick] (0.9, 0.05) -- (0.9, 0.25);
			\node[right,color=white] at (0.9, 0.25) {$y$};
		\end{tikzonimage} \\
	\end{tabular}
	\caption{\footnotesize a) HAADF-STEM image of the first frame of the image series. b) Corresponding $y$-component of peak displacements $\delta_{y}$. c) Resulting $\epsilon_{yy}$ strain map determined from the peak displacements of b).}
	\label{fig:firstFrame}
\end{figure*}

For comparison, the first frame of the image series, its determined peak displacement $\delta_{y}$ and corresponding strain map $\epsilon_{yy}$ in slow scan direction are illustrated in \cref{fig:firstFrame}. The strong artificial compression and expansion of the atomic column positions is clearly seen in the peak displacement map of \cref{fig:firstFrame} b), with a more inhomogeneous distribution in comparison to the results obtained by non-rigid registration shown in \cref{fig:kappaDisp}. This also translates into an almost uninterpretable strain map $\epsilon_{yy}$ of the single image, which is mainly dominated by slow scan distortions. In contrast, the non-rigid registration without bias correction is partly compensating the non-linear distortions and the bias-corrected approach is capable of minimizing them for the given number of frames in the series \cite{JoWeNo2017}.

\subsection{Image series for atomic scale elemental mapping}
\begin{figure}[ht]
	\centering
	\includegraphics[width=.47\textwidth]{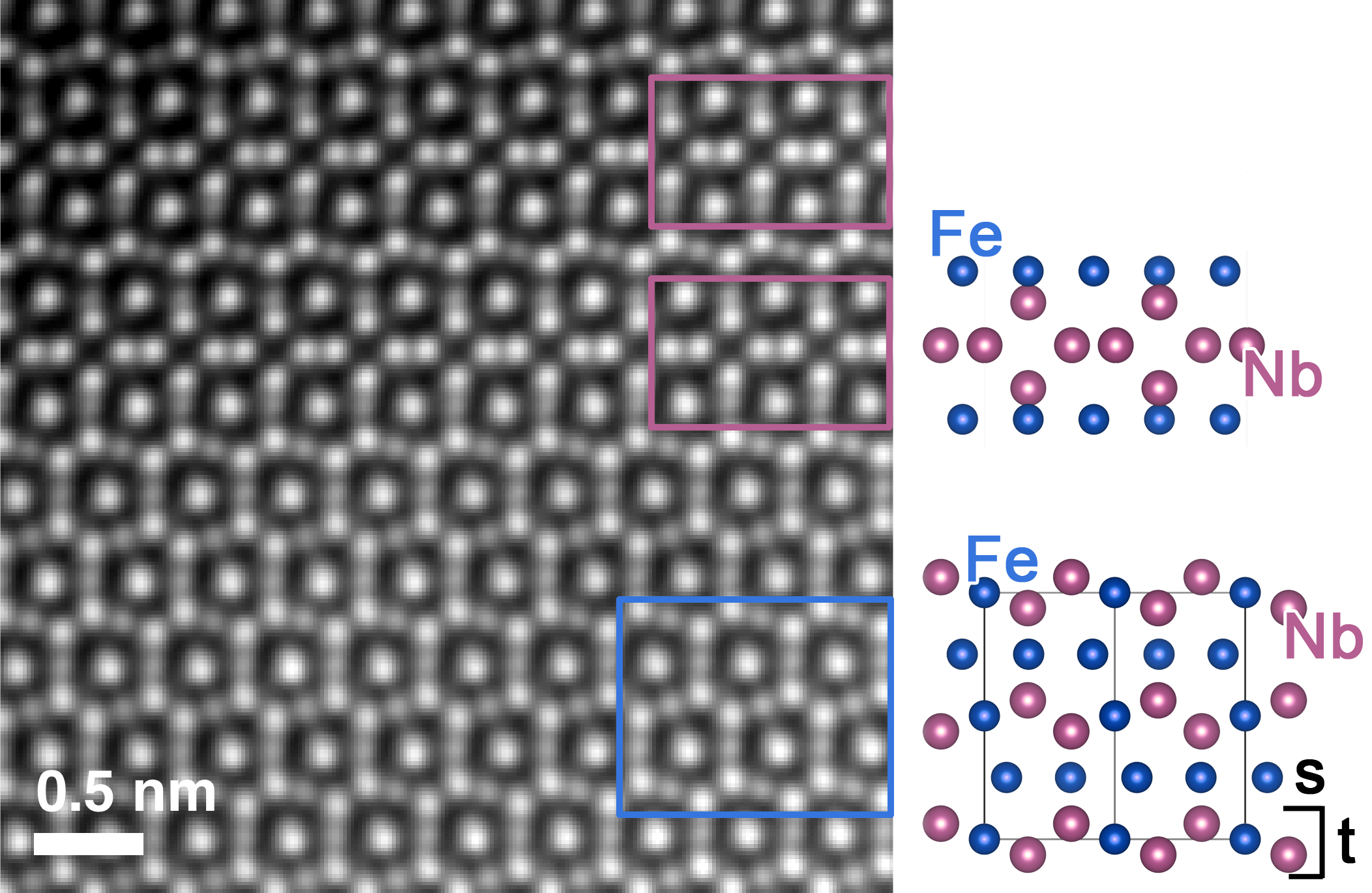}
	\caption{\footnotesize Reconstructed HAADF-STEM image of a C14 Fe$_{2}$Nb Laves phase using the bias-corrected non-rigid registration approach viewed along the $[11\overline{2}0]$ direction. The image was obtained from an image series of 240 images acquired with a pixel dwell time of 10 $\mu$s. The projected crystal structure of a region highlighted with a blue rectangle of the C14 Laves phase is shown on the right. Nb-Fe-Nb triple (t) and Fe single (s) layers are indicated. The thin black lines outline the unit cell of the C14 Fe$_{2}$Nb crystal structure. The purple rectangles highlight planar disruptions in the stacking sequence with a Nb-rich triple layer as shown in the projected crystal structure.}
	\label{fig:laves}
\end{figure}

The novel bias-corrected non-registration approach is also tested on a dataset under different experimental conditions that are optimized for elemental mapping at atomic resolution in a C14 Fe$_{2}$Nb Laves phase. The HAADF-STEM image reconstructed with the bias-corrected non-rigid registration approach viewed along the $[11\overline{2}0]$ direction and projected crystal structures are shown in \cref{fig:laves}. Here, a series of 240 images with a pixel dwell time of 10 $\mu$s is aligned and registered under typical conditions used for atomic resolution elemental mapping.

\begin{figure*}[ht]
	\setlength{\imgwidth}{0.23\linewidth}
	\centering
	\setlength{\tabcolsep}{0.75ex}
  \begin{tabular}{ccc}
    &
  	&
  	Peak position offset [pixels]\\
	\resizebox{\imgwidth}{!}{
	\begin{tikzpicture}
	\begin{axis}[
	    hide axis,
	    scale only axis,
	    height=0pt,
	    width=0pt,
			xmin=0,xmax=1,
			ymin=0,ymax=1,
	    colormap/thermal,
	    colorbar horizontal,
	    point meta min=-0.07,
	    point meta max=0.05,
	    colorbar style={
	        width=5cm,
					height=0.25cm,
					xtick distance=0.03,
					scaled ticks=false,
					xticklabel style={/pgf/number format/fixed}
	    }]
	    \addplot [draw=none] coordinates {(0,0)};
	\end{axis}
	\end{tikzpicture}}&
	\resizebox{\imgwidth}{!}{
	\begin{tikzpicture}
	\begin{axis}[
			hide axis,
			scale only axis,
			height=0pt,
			width=0pt,
			xmin=0,xmax=1,
			ymin=0,ymax=1,
			colormap/thermal,
			colorbar horizontal,
			point meta min=-0.07,
			point meta max=0.05,
			colorbar style={
					width=5cm,
					height=0.25cm,
					xtick distance=0.03,
					scaled ticks=false,
					xticklabel style={/pgf/number format/fixed}
			}]
			\addplot [draw=none] coordinates {(0,0)};
	\end{axis}
	\end{tikzpicture}}&
	\resizebox{\imgwidth}{!}{
	\begin{tikzpicture}
	\begin{axis}[
			hide axis,
			scale only axis,
			height=0pt,
			width=0pt,
			xmin=0,xmax=1,
			ymin=0,ymax=1,
			colormap/jet,
			colorbar horizontal,
			point meta min=0.08,
			point meta max=0.46,
			colorbar style={
					width=5cm,
					height=0.25cm,
					xtick distance=0.1,
					scaled ticks=false,
					xticklabel style={/pgf/number format/fixed}
			}]
			\addplot [draw=none] coordinates {(0,0)};
	\end{axis}
	\end{tikzpicture}}\\
	\begin{tikzonimage}[width=\imgwidth]{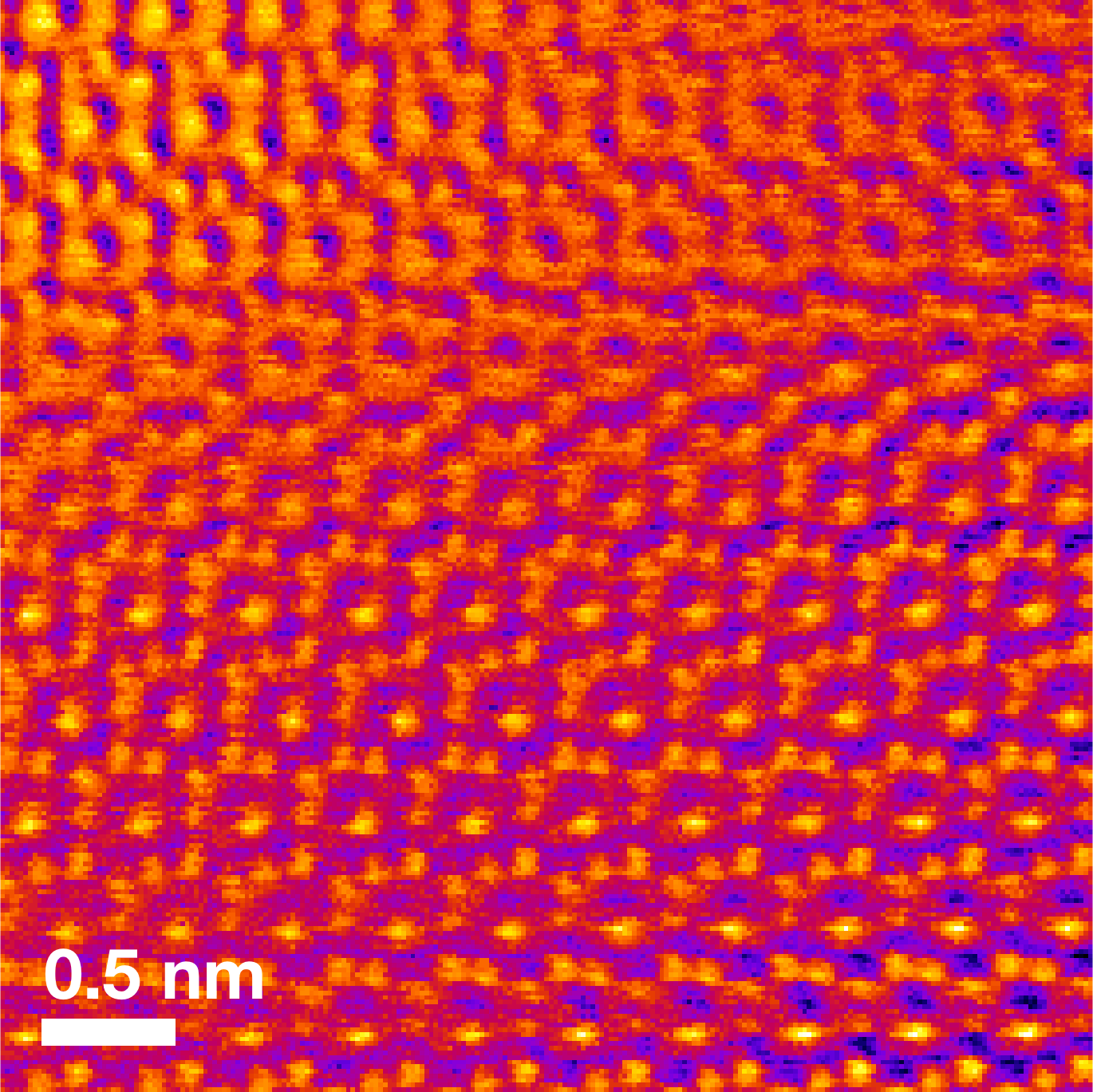}
	\node[color=white] at (0.1,0.9) {\Large a)};
  \end{tikzonimage}&
	\begin{tikzonimage}[width=\imgwidth]{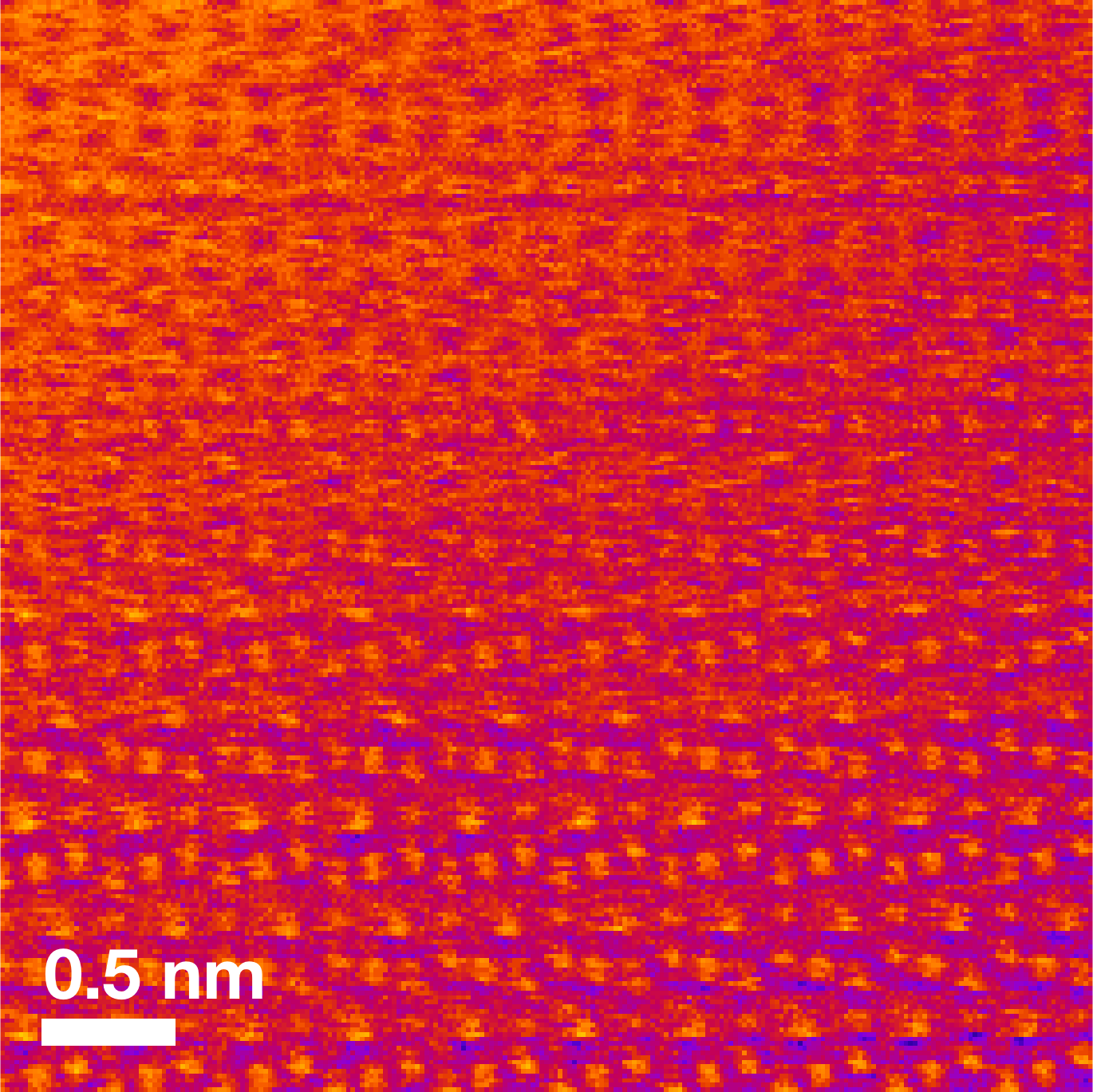}
	\node[color=white] at (0.1,0.9) {\Large b)};
  \end{tikzonimage}&
	\begin{tikzonimage}[width=\imgwidth]{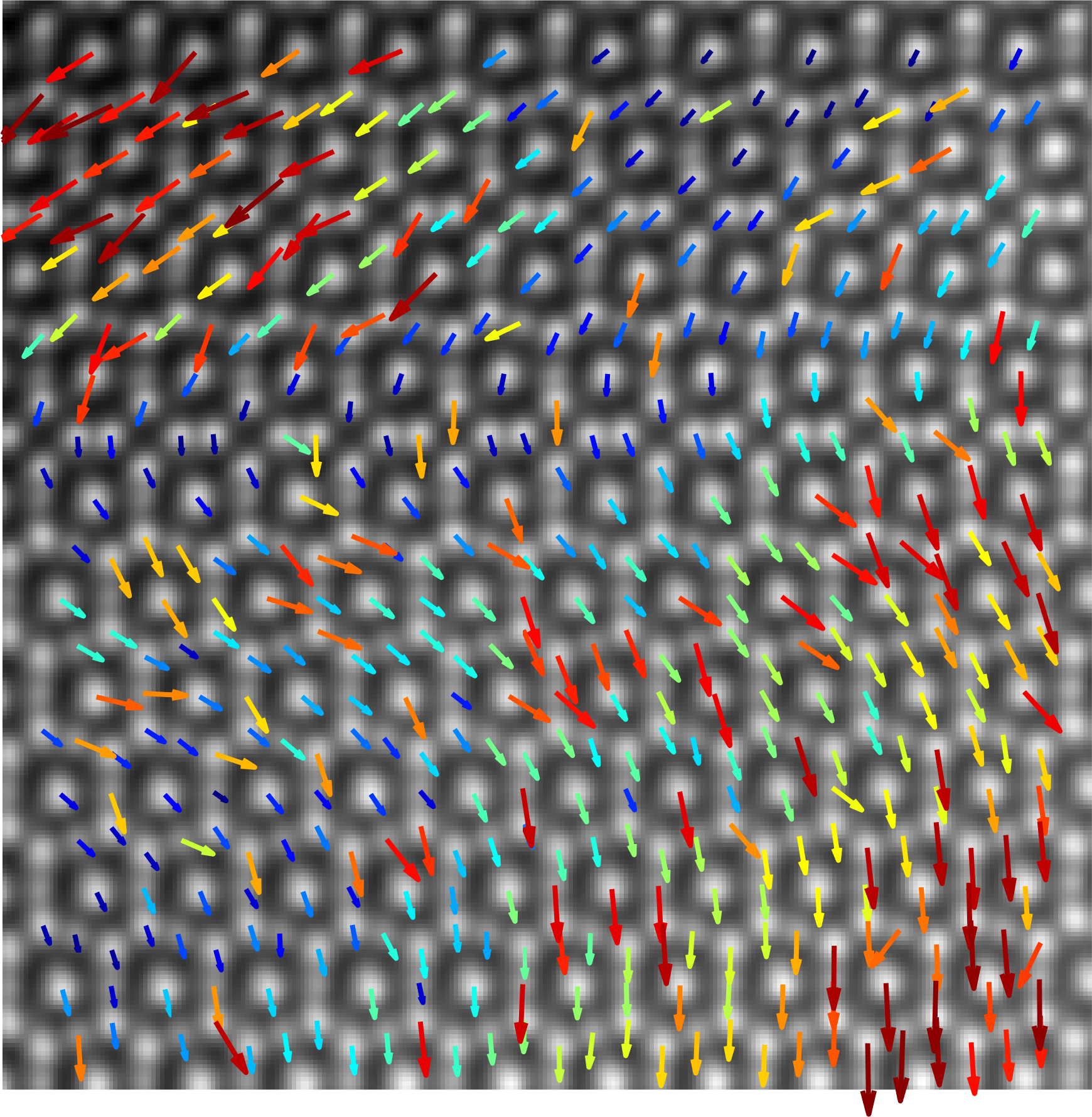}
	\node[color=white] at (0.1,0.9) {\Large c)};
  \end{tikzonimage}\\
	$f_\text{NR,no corr}-f_\text{R}$&
	$f_\text{NR,corr}-f_\text{R}$
	\end{tabular}
	\caption{\footnotesize Normalized difference images for a) non-rigid without bias-correction minus rigid, b) non-rigid with bias-correction minus rigid. c) Vector field of the difference vectors determined from fitted lattice peak positions of the bias-corrected and non-bias corrected non-rigid registration approach.}
	\label{fig:lavesDiff}
\end{figure*}

The atomic structure of the Laves phase in this viewing direction can be described as a layered arrangement of alternating triple and single layers of atoms. The triple layers, containing both Fe and Nb atoms are separated by a single layer of Fe atoms arranged in a kagom\'{e} net \cite{Chisholm2005a} as highlighted in the projected crystal structure of \cref{fig:laves}. Comparing the difference images obtained by the non-rigid registration approaches with respect to the series aligned with rigid registration reveals that local distortions impact atomic column peak positions, as seen in Figs. \ref{fig:lavesDiff} a) and b). A strong influence is readily visible in the difference image of the non-rigid registration result without bias-correction and the rigidly aligned image series shown in \cref{fig:lavesDiff} a). From the top left to the bottom right of the image a continuous distortion of atom column positions is found. The residual features in the difference image of the non-rigid-registration result with bias-correction and the mean image of the rigid series are less pronounced as seen in \cref{fig:lavesDiff} b). The strongest difference is seen in the lower right corner of the difference image and in the rest only residual contributions from fast scan noise can be seen as small horizontal stripes. The vector field of \cref{fig:lavesDiff} c) shows the vector difference of the lattice peak positions of the bias-corrected and non-corrected non-rigid data sets superimposed on the mean image of the non-rigidly aligned series without bias correction. The atomic column positions were obtained by peak fitting with 2D non-linear Gaussian functions. The maximum peak location offset of the non- and bias-corrected approaches is 0.5 pixel and the vectors are scaled for visualization in \cref{fig:lavesDiff} c). The strongest deviation is seen in the upper left and lower right corner of the vector field of \cref{fig:lavesDiff} c). However, the small magnitude in peak position offset indicates that both methods are capable of reducing the effect of non-linear distortions in the current dataset.

\begin{figure*}[ht]
	\setlength{\imgwidth}{0.23\linewidth}
	\centering
	\setlength{\tabcolsep}{0.75ex}
	\begin{tabular}{ccc}
		\resizebox{\imgwidth}{!}{
			\begin{tikzpicture}
			\begin{axis}[
			hide axis,
			scale only axis,
			height=0pt,
			width=0pt,
			xmin=0,xmax=1,
			ymin=0,ymax=1,
			colormap/jet,
			colorbar horizontal,
			point meta min=-1,
			point meta max=1,
			colorbar style={
				width=5cm,
				height=0.25cm,
				xtick distance=.5,
				scaled ticks=false,
				xticklabel style={/pgf/number format/fixed}
			}]
			\addplot [draw=none] coordinates {(0,0)};
			\end{axis}
			\end{tikzpicture}\raisebox{1ex}{[\%]}
		}&
		\resizebox{\imgwidth}{!}{
			\begin{tikzpicture}
			\begin{axis}[
			hide axis,
			scale only axis,
			height=0pt,
			width=0pt,
			xmin=0,xmax=1,
			ymin=0,ymax=1,
			colormap/jet,
			colorbar horizontal,
			point meta min=-1,
			point meta max=1,
			colorbar style={
				width=5cm,
				height=0.25cm,
				xtick distance=.5,
				scaled ticks=false,
				xticklabel style={/pgf/number format/fixed}
			}]
			\addplot [draw=none] coordinates {(0,0)};
			\end{axis}
			\end{tikzpicture}\raisebox{1ex}{[\%]}
		}&
		\resizebox{\imgwidth}{!}{
			\begin{tikzpicture}
			\begin{axis}[
			hide axis,
			scale only axis,
			height=0pt,
			width=0pt,
			xmin=0,xmax=1,
			ymin=0,ymax=1,
			colormap/jet,
			colorbar horizontal,
			point meta min=-1,
			point meta max=1,
			colorbar style={
				width=5cm,
				height=0.25cm,
				xtick distance=.5,
				scaled ticks=false,
				xticklabel style={/pgf/number format/fixed}
			}]
			\addplot [draw=none] coordinates {(0,0)};
			\end{axis}
			\end{tikzpicture}\raisebox{1ex}{[\%]}
		}\\
		\begin{tikzonimage}[width=\imgwidth]{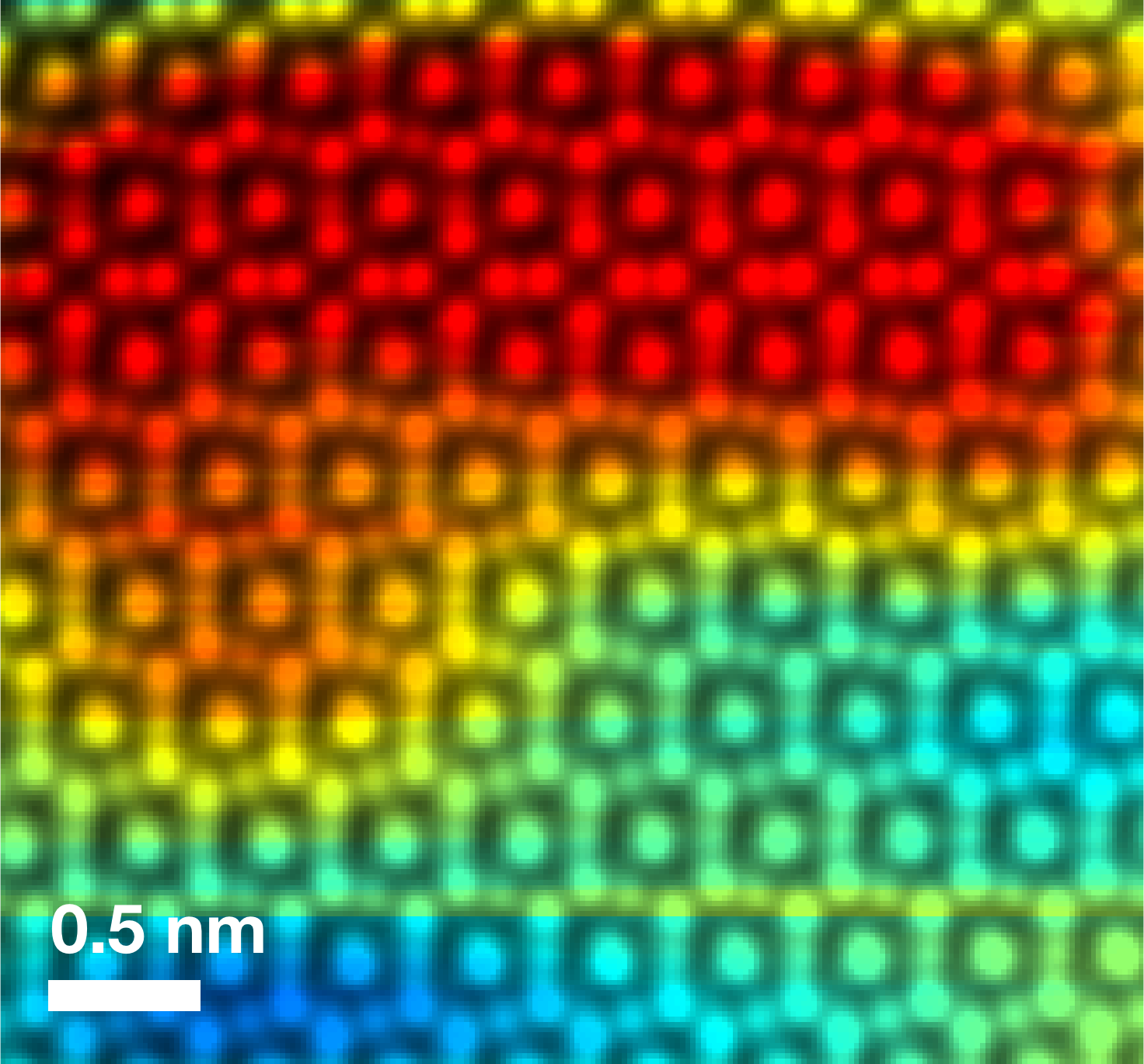}
			\node[color=white] at (0.1,0.9) {\Large a)};
			\node[color=white] at (0.9,0.9) {\Large $\epsilon_{yy}$};
			\draw[->,color=white,thick] (0.9, 0.05) -- (0.9, 0.25);
			\node[right,color=white] at (0.9, 0.25) {$y$};
			\draw[->,color=black,ultra thick] (0.4, 0.25) -- (0.2, 0.375);
		 	\draw[->,color=black,ultra thick] (0.5, 0.2) -- (0.3, 0.1);
		 	\draw[dotted,color=white,ultra thick] (0,0.55) -- (1,0.55);
			\node[right,color=white] at (0.5, 0.1) {\Large C14};
			\node[right,color=white] at (0.3, 0.9) {\Large defect};
		\end{tikzonimage}&
		\begin{tikzonimage}[width=\imgwidth]{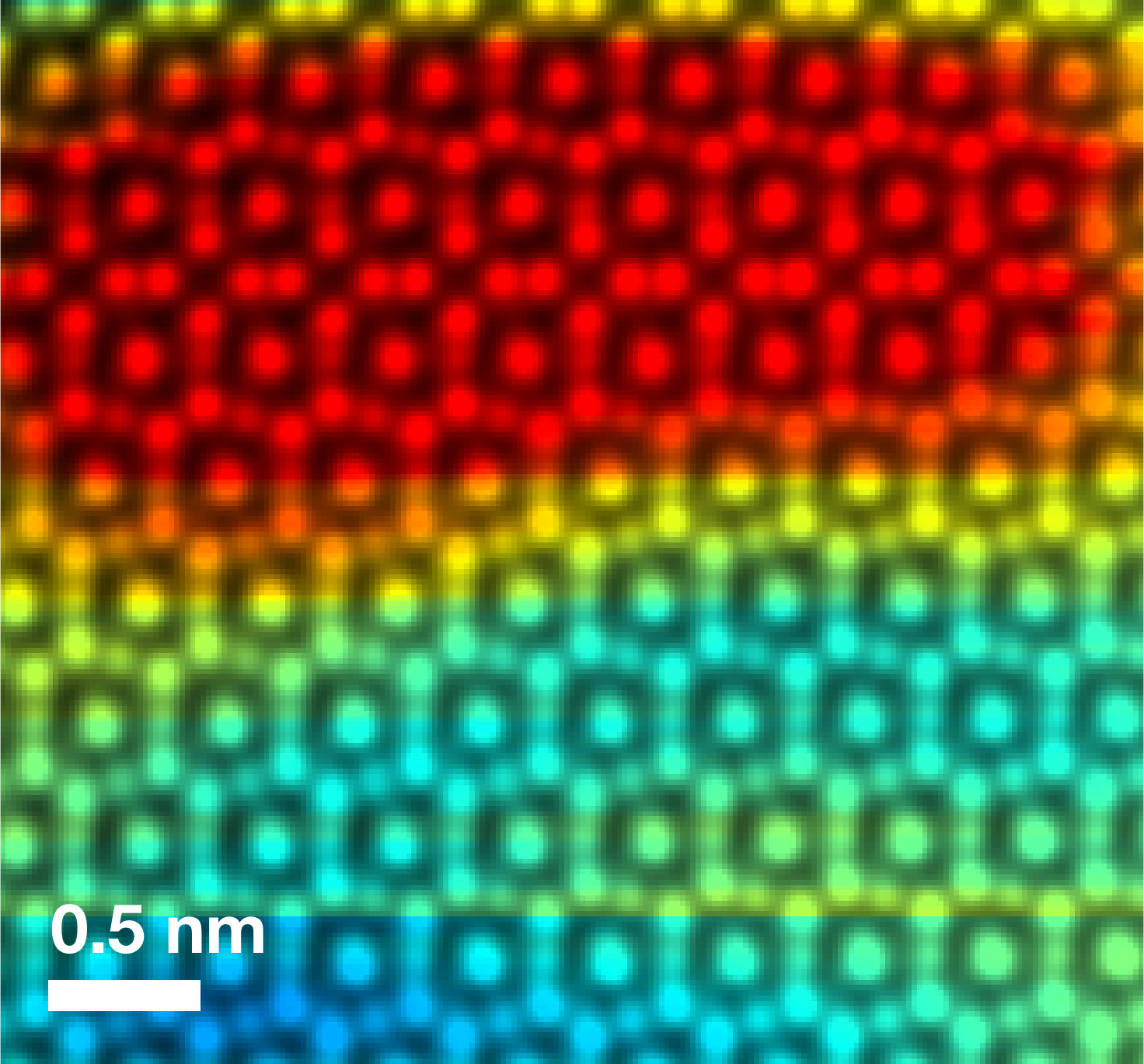}
			\node[color=white] at (0.1,0.9) {\Large b)};
			\node[color=white] at (0.9,0.9) {\Large $\epsilon_{yy}$};
			\draw[->,color=white,thick] (0.9, 0.05) -- (0.9, 0.25);
			\node[right,color=white] at (0.9, 0.25) {$y$};
			\draw[->,color=black,ultra thick] (0.4, 0.25) -- (0.2, 0.375);
			\draw[->,color=black,ultra thick] (0.5, 0.2) -- (0.3, 0.1);
			\draw[dotted,color=white,ultra thick] (0,0.55) -- (1,0.55);
		\end{tikzonimage}&
		\begin{tikzonimage}[width=\imgwidth]{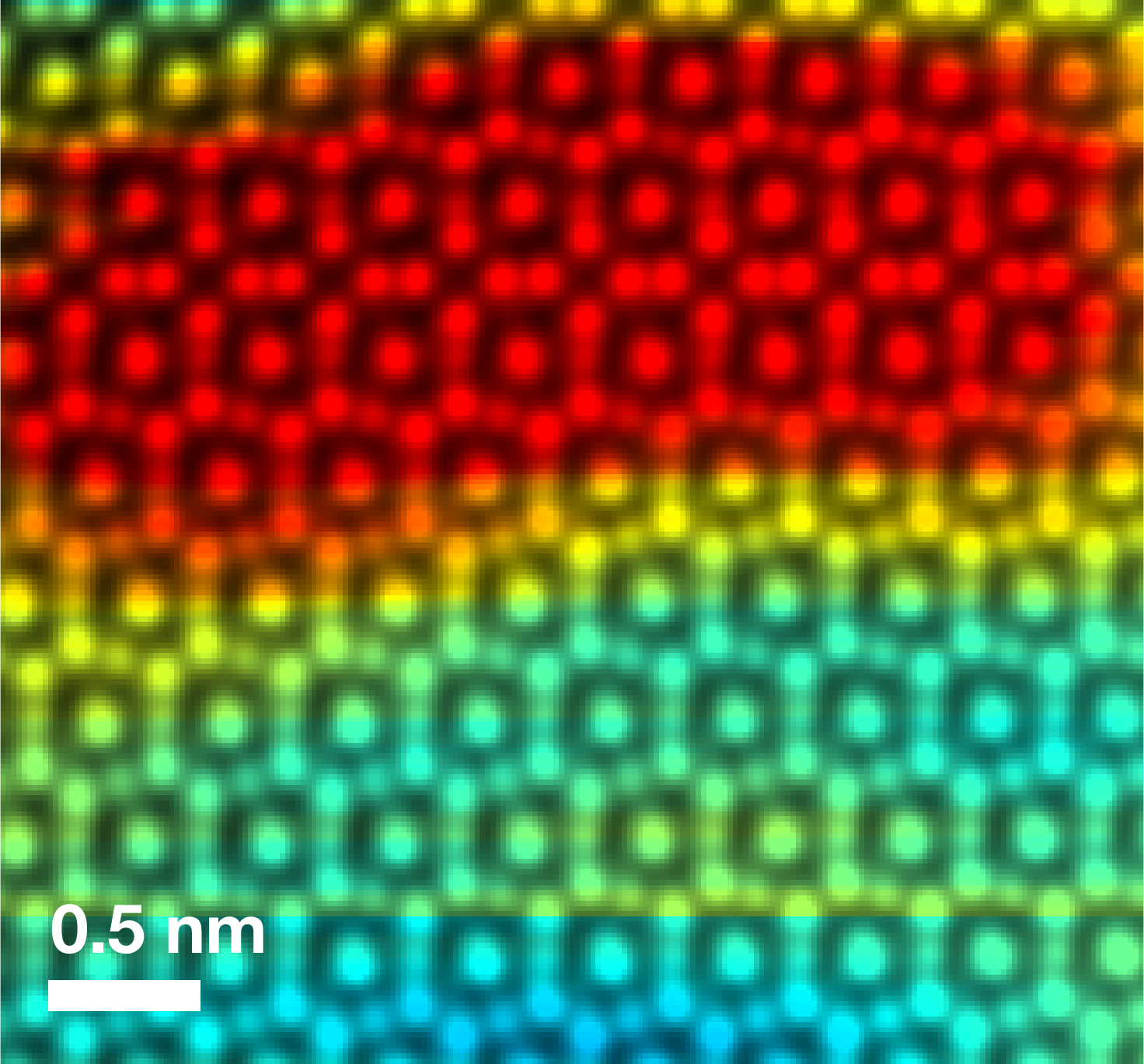}
			\node[color=white] at (0.1,0.9) {\Large c)};
			\node[color=white] at (0.9,0.9) {\Large $\epsilon_{yy}$};
			\draw[->,color=white,thick] (0.9, 0.05) -- (0.9, 0.25);
			\node[right,color=white] at (0.9, 0.25) {$y$};
			\draw[->,color=black,ultra thick] (0.4, 0.25) -- (0.2, 0.375);
			\draw[->,color=black,ultra thick] (0.5, 0.2) -- (0.3, 0.1);
			\draw[dotted,color=white,ultra thick] (0,0.55) -- (1,0.55);
		\end{tikzonimage} 	 \\
	rigid&
	no bias-correction&
	with bias-correction \\
	\end{tabular}
	\caption{\footnotesize $\epsilon_{yy}$ strain components determined from the fitted atomic column positions of a) the rigidly aligned dataset, b) the data set aligned by non-rigid registration without bias correction and c) the non-rigid one with bias correction. The black arrows indicate regions with the most pronounced difference between the rigid and non-rigid datasets. The white horizontal line is a guide for the eyes separating the C14 Fe$_{2}$Nb phase from the one containing planar defects.}
	\label{fig:lavesStrain}
\end{figure*}

For a better comparison of the effect between the different image registration procedures, the $\epsilon_{yy}$ strain components determined from the fitted lattice peak positions are illustrated in \cref{fig:lavesStrain}. The area in the images under tensile strain, highlighted in red, represents regions of the structure containing planar defects. They cause a disruption in the stacking sequence of the C14 Laves phase, as indicated in \cref{fig:laves}, and correspondingly an expansion of the lattice. The black arrows highlight areas in the strain maps where an improvement is observed by employing the non-rigid registration approaches. Since the reference lattice was taken in the C14 Fe$_{2}$Nb phase, the areas under slight compression and expansion in \cref{fig:lavesStrain} a) can be related to residual non-linear distortions in the rigidly aligned image. These effects are best compensated in the bias-corrected dataset shown in \cref{fig:lavesStrain} c), but the difference to the non-bias corrected approach of \cref{fig:lavesStrain} b) is negligible.

\section{Discussion}
The ability to image complex materials and sample geometries in a STEM opened manifold possibilities to explore composition, strain and chemical bonding in materials at atomic resolution. A crucial aspect to enable quantitative imaging in STEM is to reduce scan noise artifacts \cite{KiAsYu10,BrBoLa2012,JoNe13} that otherwise obstruct and complicate the precision to locate the position of atomic columns. Newly developed image alignment techniques are able to measure and compensate non-local distortions in image series and with this are effective in reducing scan noise \cite{Berkels2012,Berkels2012,JoWeNo2017}. However, scan noise and instabilities are strongly depending on the microscope and room conditions. Hence, flexible and adaptable algorithms are required to compensate for the influence of instabilities in the most effective way.

The extension of the non-rigid registration approach developed in \cite{BeBiBl13} presented here is demonstrated on a synthetic dataset containing both fast and slow scan noise. The biased-corrected approach is able to compensate for the non-local distortions, since the deformations are no longer biased towards the first frame in the series, but to the average coordinate system of the input series. This is achieved by removing systematic differences of the individual deformations to the identity mapping in an additional step. The bias-corrected approach is effectively reducing the non-local distortions stemming from artificial slow scan noise, as well as fast scan noise artifacts.

We also test how the bias correction in the new approach affects the determination of strain fields from atomic resolution STEM images. Here, we are focusing on the component in slow scan direction and it becomes evident that the influence of slow scan noise is reduced by more than 50$\%$ considering the total spread from negative to positive strain values as illustrated in \cref{fig:kappaStrain}. The mean and standard deviation of the $\epsilon_{yy}$ component are reduced from 2.0$\%$ $\pm$ 0.13$\%$ to 0.2$\%$ $\pm$ 0.13$\%$ for the approaches without and with bias-correction, respectively. As already pointed out by Jones et al.\ \cite{Jones2018}, the accuracy of peak fitting is another limiting factor in determining precise strain values, which is largely affected by the sample quality and the number of images in a dataset \cite{JoWeNo2017}. In comparison, a single frame of the image series is largely dominated by slow scan noise making the data almost uninterpretable. According to the 'distortion-relative to a single frame' criterion proposed by Jones et al.~\cite{JoWeNo2017}, the approach without bias correction reduces slow scan noise by 21 $\%$ and the bias-corrected non-rigid registration by 48 $\%$ with respect to the single frame.

In our case, the measurements were taken on the apex of a needle-shaped specimen prepared by focused ion beam milling, which introduces surface damage that obstructs the precise determination of lattice peak positions. The strain measurement suggests that even with the bias-corrected approach it is difficult to fully correct for the influence of slow scan instabilities, as also observed by \cite{Jones2018}. The approach developed by Ophus et al.\ \cite{Ophus2016a} only relying on two orthogonally scanned images is also able to greatly reduce the influence of scan noise in comparison to individual slow scan images, but also in their case, residual non-local distortions remain.

The novel non-rigid registration approach is also applicable to different experimental conditions, as shown in \cref{fig:laves,fig:lavesDiff}. The image series here is taken on the same microscope, but with larger pixel dwell time and more than 200 images in the series, representing conditions for atomic resolution chemical mapping. In this image series, the influence of slow scan noise is of minor impact, but still an inhomogeneous distortion exists in the images reconstructed by rigid alignment, since it only accounts for image translations in the series. The first image contains large non-local distortions and hence these deformations are transferred to the average of the entire image series by the bias-uncorrected non-rigid registration, leading to an overestimation of the average deformations. This also explains the strong deviations in the difference image of \cref{fig:lavesDiff} c). Non-rigid registration has been successfully applied to improve the influence of scan distortions on atomically resolved elemental maps determined by energy dispersive X-ray spectroscopy (EDS) \cite{Jones2018,YaZhOh16}. Since spectroscopic imaging typically requires the collection of the X-ray signal over several minutes, the sample stability and experimental conditions are even more decisive to obtain interpretable results. The comparison of the rigid registration with the bias-corrected non-rigid alignment of \cref{fig:lavesDiff} b) shows that especially in the lower right corner of the image, atomic column positions are displaced, which is not accounted for by the rigid alignment. This certainly influences not only the quality and distortions in the reconstructed image, but will also affect the quality of spectroscopic elemental maps \cite{Jones2018}. Both presented non-rigid registration techniques are capable of minimizing these residual non-local distortions and produce similar results with a maximum peak displacement offset of 0.5 pixels. The comparison of the $\epsilon_{yy}$ strain maps illustrates that slow scan instabilities are less affecting the overall measurement quality, but non-local distortions are still present in the rigidly aligned dataset. This effect can be related to the different experimental conditions, compared to the previous example, since the series here was taken at higher magnification and hence the electron beam is moving over a smaller area as well as the largely increased number of frames in the series \cite{JoWeNo2017}.

Instabilities in STEM are depending on the local room conditions, microscope stability and can vary on a daily basis. The presented non-rigid registration can be used to monitor these instabilities and is adaptable to different experimental conditions, either used for atomic resolution strain or elemental mapping. With the developed methodology, it is even possible to determine the characteristics of slow scan effects so that the experimental conditions can be optimized to minimize their impact on the experiment. Residual distortions can then be compensated by the bias-corrected non-rigid registration to obtain images with a minimum amount of distortions.

\section{Conclusions}
To conclude, a novel extension of a non-rigid image registration is developed to reduce non-local distortions in STEM image series. The approach is versatile and can be applied to different serial imaging conditions whenever instrumental instabilities are degrading image quality in a STEM. The bias-corrected non-rigid registration is able to compensate for slow scan artifacts that otherwise obstruct the precise determination of atomically resolved strain fields within a sample. The approach is also applicable to long exposures at high pixel dwell times and several hundreds of frames per series. The presented methodology is very robust even when strong distortions are present and vary within the STEM image series.

It is envisioned that the new approach will also be coupled with spectroscopic signals and advanced denoising techniques to improve the quality and interpretability of atomic resolution elemental maps. The algorithm could also be directly used at the microscope to monitor the noise performance of the microscope and to optimize the image acquisition accordingly.

The {C++} source code of the proposed method is available at \url{https://github.com/berkels/match-series}.

\section*{Acknowledgements}
B.\ Berkels was funded in part by the Excellence Initiative of the German Federal and State Governments. The authors kindly acknowledge Dr.\ Mengji Yao (MPIE) for preparing the low-density steel samples. We also thank Dr.\ Frank Stein (MPIE) and Dr.\ Michaela Šlapáková (Charles University, Prague) for providing the Fe$_2$Nb alloy.

\bibliographystyle{unsrt}
\bibliography{refs_final}

\end{document}